\documentclass[12pt]{article}
\usepackage{graphicx}
\usepackage{authblk}
\usepackage{orcidlink,amsmath,rotating,amssymb,amsthm,algorithm,epstopdf,multicol,multirow,gensymb,subcaption,fontenc,algorithmic,tablefootnote,longtable,array}
\usepackage[section]{placeins}
\usepackage{enumerate}
\usepackage{natbib}
\usepackage{url} 
\usepackage{hyperref}
\usepackage{xcolor,soul}
\hypersetup{colorlinks=true,linkcolor=,citecolor=blue}

\newcommand{\blind}{1}

\begin{document}

\def\spacingset#1{\renewcommand{\baselinestretch}%
{#1}\small\normalsize} \spacingset{1}


\if1\blind
{
  \title{\bf Model-based clustering using a new mixture of circular regressions}
  \author{Sphiwe Bonakele Skhosana\textsuperscript{*}\orcidlink{0000-0002-4740-7540} and Najmeh Nakhaei Rad\orcidlink{0000-0002-7831-5614}\vspace{.5cm}\\
    Department of Statistics, University of Pretoria, Pretoria, 0028, South Africa\vspace{.5cm}\\
    \textsuperscript{*}Corresponding author: \url{spiwe.skhosana@up.ac.za}}
  \maketitle
} \fi

\if0\blind
{
  \bigskip
  \bigskip
  \bigskip
  \begin{center}
    {\LARGE\bf Mixtures of circular regressions}
\end{center}
  \medskip
} \fi

\bigskip
\begin{abstract}
Regression models, where the response variable is circular, are common in areas such as biology, geology and meteorology. A typical model assumes that the conditional distribution of the response follows a von-Mises distribution. However, this assumption is inadequate when the response variable is multimodal. For this reason, in this paper, a finite mixture of regressions model is proposed for the case of a circular response variable and a set of circular and/or linear covariates. Mixture models are very useful when the underlying population is multimodal. Despite the prevalence of multimodality in regression modelling of circular data, the use of mixtures of regressions has received no attention in the literature. This paper aims to close this knowledge gap. To estimate the proposed model, we develop a maximum likelihood estimation procedure via the Expectation-Maximization algorithm. An extensive simulation study is used to demonstrate the practical use and performance of the proposed model and estimation procedure. In addition, the model is shown to be useful as a model-based clustering tool. Lastly, the model is applied to a real dataset from a wind farm in South Africa.
\end{abstract}

\noindent%
\vfill

\newpage
\spacingset{1.4} 

\section{Introduction}\label{sec:intro}
Data with “directional” attributes (angles) known as circular data is considered unusual and cannot be modelled using Cartesian coordinate systems due to their periodic nature. Consequently, there is a necessity to develop statistical techniques beyond the conventional analytical frameworks to effectively model, interpret, and utilize such data across diverse scientific domains. A circular-linear regression model arises when the response variable, denoted by $\theta$, is measured on a circle and the covariates $\mathbf{x}\in \mathbf{R}^p$ are measured on the real line. Such situations are commonly encountered in areas such as biology, geology and meteorology, among many other scientific fields. For conservation purposes, it may be important to, for example, study the dependence of the direction an animal moves on the distance moved (\cite{fisher1992}) or study the dependence of the spawning time of a fish on, say the amplitude of low-tide (\cite{lund1999}). For applications in geology and meteorology, see the examples mentioned in \cite{fisher1992} and \cite{jammalamadaka2001}.\\
The circular-linear regression model was first developed by \cite{gould1969} through assuming that the error distribution has a von-Mises distribution. \cite{johnson1978} showed that the Gould model has a serious problem in that the regression function spirals infinitely many times around a cylinder similar to a rotating "barber's pole". This implies that the likelihood function of the model has infinitely many maxima of equal importance leading to a model that is not identifiable and hence not useful in practice. For a single covariate, the authors (\cite{johnson1978}) improved the Gould model by proposing a model in which the regression function completes a spiral around a cylinder resulting in a unique model with a unique solution. \cite{fisher1992} generalised the Johnson-Wehrly model to accommodate multiple covariates and further extended the model for the case in which the dispersion of the circular response depends on the covarites. For more details about the circular-linear regression model and, in general, regression models for circular or directional data, see (\cite{fisher1992}; sec. 6.4) and (\cite{mardia2000}; sec. 11.3). \\
More and more data sets on the circle tend to show non-trivial features such as multimodality (\cite{ley2017modern}). The circular-linear regression model is not an adequate model when the circular response $\theta$ is multimodal, which is common in practice. For instance, in the above animal orientation example, animals may have to choose between, say two directions of movement both of which are more orless preferred \cite{batschelet1981}. In a 24-hour cycle, the peak spawning hour of a certain species of fish may occur in the early hours of the evening (say, around 6pm) or midnight (say, around 12am) (\cite{child1991}). Sample data that exhibit such patterns may be said to have been generated by two (or more) unimodal distributions. Unlike the examples given above, for most cases in practice, the angle (or time) between the multiple directions (or time-peaks) may be arbitrary with no standard way of separating the observed data into unimodal data samples (\cite{batschelet1981}). An objective way to handle multimodality in circular or, in general, directional data analysis is to make use of mixture models (\cite{fisher1993} and \cite{mclachlan2000}). \\
Mixture models are useful for modelling sample data that come from a population that is made up of, say $K$, unknown sub-populations, in which case they are called finite mixture models (\cite{fruhwirth2006}). Each sub-population can have its own unique probability distribution. Earlier uses of finite mixture models include Newcomb (\cite{newcomb1886}) and Pearson (\cite{pearson1894}) in astronomy and zoology, respectively. Finite mixture models were later extended to areas such as finance for modelling financial returns (\cite{kon1984}), economics for modelling income distribution (\cite{schneider2020}), ecology for modelling species distribution (\cite{joseph2009}), among other areas of application. Applications of finite mixture models for circular data include \cite{stephens1969} who, amongst other things, fitted a $K=2$ component mixture of von-Mises distribution to data on turtle movement. The mixture of von-Mises distributions has since been used for modelling seasonality in sudden infant death syndrome (\cite{mooney2003}), for modelling wind direction (\cite{carta2008}) and \cite{masseran2013}) and for bearing-only tracking (\cite{markovic2012}) in robotics. Recently, a mixture of von-Mises distribution was proposed for clustering circular data (\cite{jammalamadaka2024}). To account for outliers in the data, a mixture of a von-Mises distribution and a uniform distribution was proposed in \cite{bentley2006}. To account for skewness in the data, \cite{rad2022} proposed a mixture of sine-skewed von-Mises distributions to model wind direction.\\
A highly useful class of the general finite mixture model is the finite mixture of regressions model. Due to their flexibility, finite mixtures of regressions allow each sub-population to have its own unique relationship between a response variable and a set of covariates. Finite mixtures of regressions have their origin in economics as switching regression models (\cite{quandt1972}, \cite{goldfeld1973} and \cite{quandt1978}). Their success in economics has led to widespread adoption in areas such as marketing (\cite{desarbo1988} as latent class regression models), machine learning (\cite{jacobs1991} as mixtures-of-experts models) and biology, among many other areas of application. In these areas of application, finite mixtures of regressions underpin a variety of statistical modelling techniques such as robust regression modelling (\cite{box1968}), model-based clustering (\cite{ingrassia2014}), distributional regression (\cite{kneib2023}), mixed-effects modelling (\cite{fruhwirth2006}; sec. 8.5) and censored data modelling (\cite{zeller2019}). Comprehensive surveys of finite mixtures of regressions and finite mixture models, in general, can be found in (chapter 8 of \cite{fruhwirth2006} and chapter 3 of \cite{yao2024}) and (\cite{mclachlan2000} and \cite{fruhwirth2019}), respectively.\\
Unfortunately, the use of finite mixtures of regressions for circular data has received no attention in the literature. Thus, there exists a knowledge gap, the closure of which enriches the literature on finite mixture modelling and circular data analysis. In this paper, we propose a finite mixture of regressions model, henceforth referred to as a finite mixture of circular regressions model.
\begin{figure}[h]
   \centering
    \begin{subfigure}[b]{0.45\textwidth}
   \includegraphics[width=\textwidth]{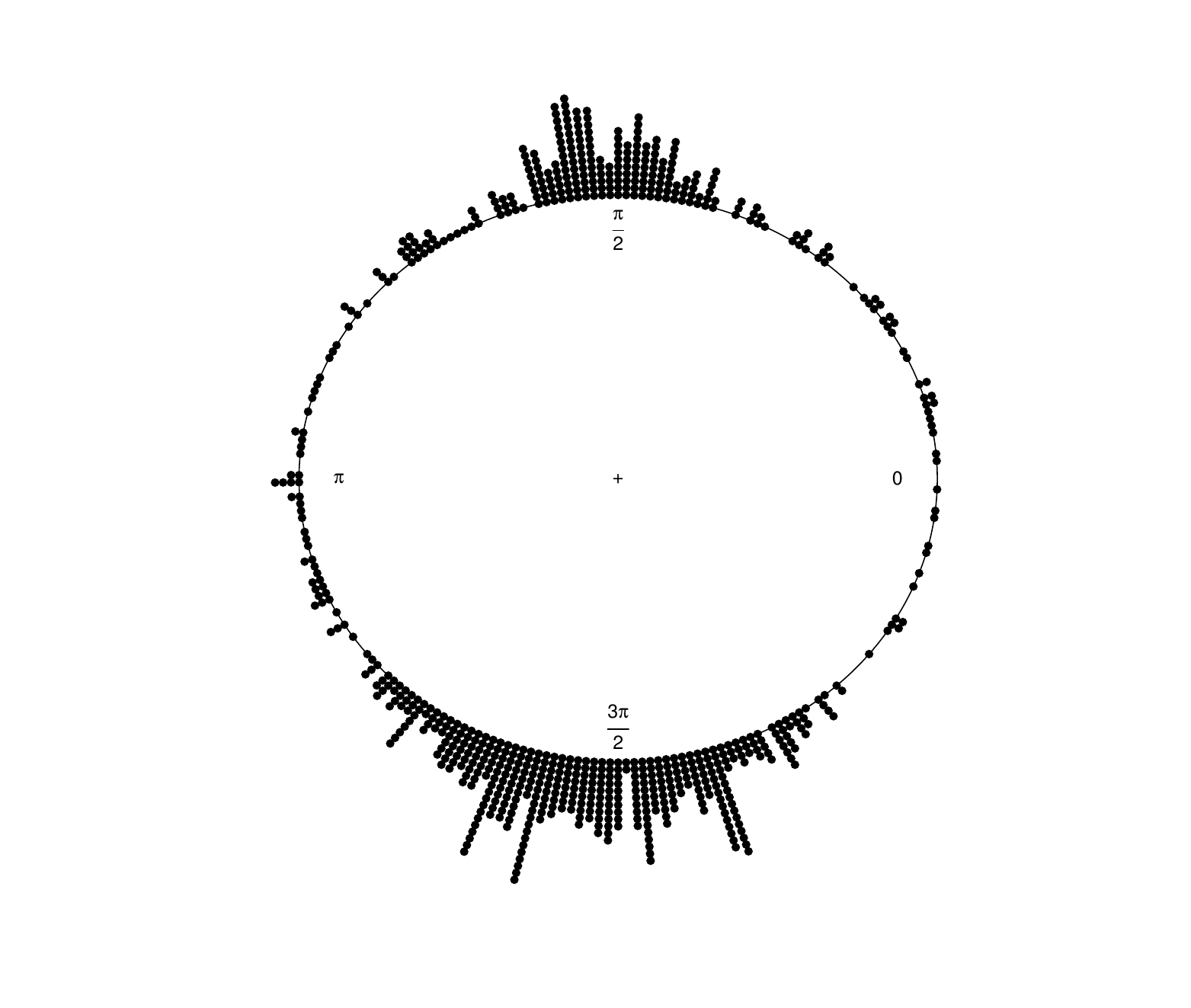}
    \caption{}
    \label{fig:circ_wind}
    \end{subfigure}
    \hfill
    \begin{subfigure}[b]{0.45\textwidth}
    \includegraphics[width=\textwidth]{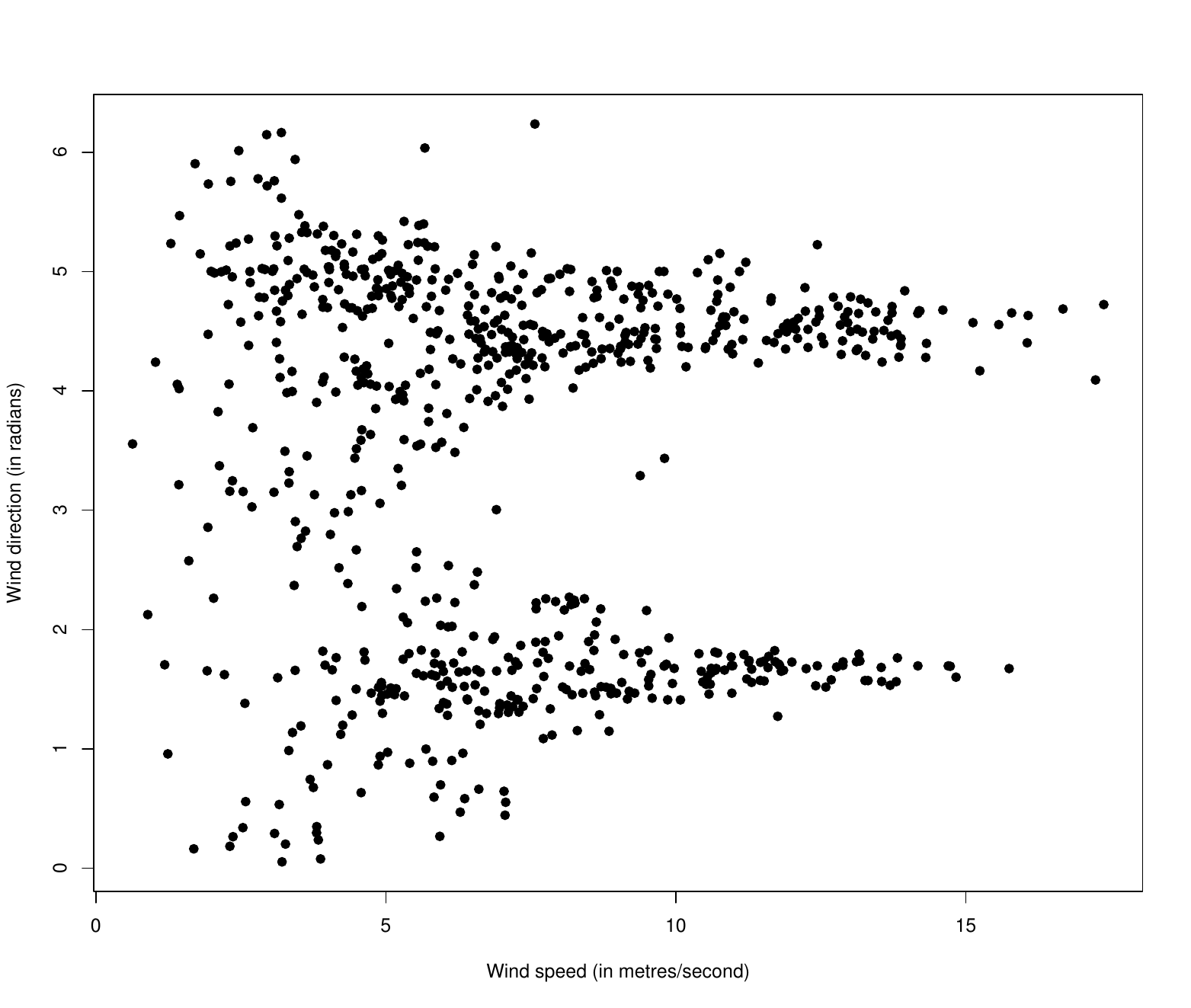}
    \caption{}
    \label{fig:scatter_windDS}
    \end{subfigure}
    \vfill
    \begin{subfigure}[b]{0.45\textwidth}
    \includegraphics[width=\textwidth]{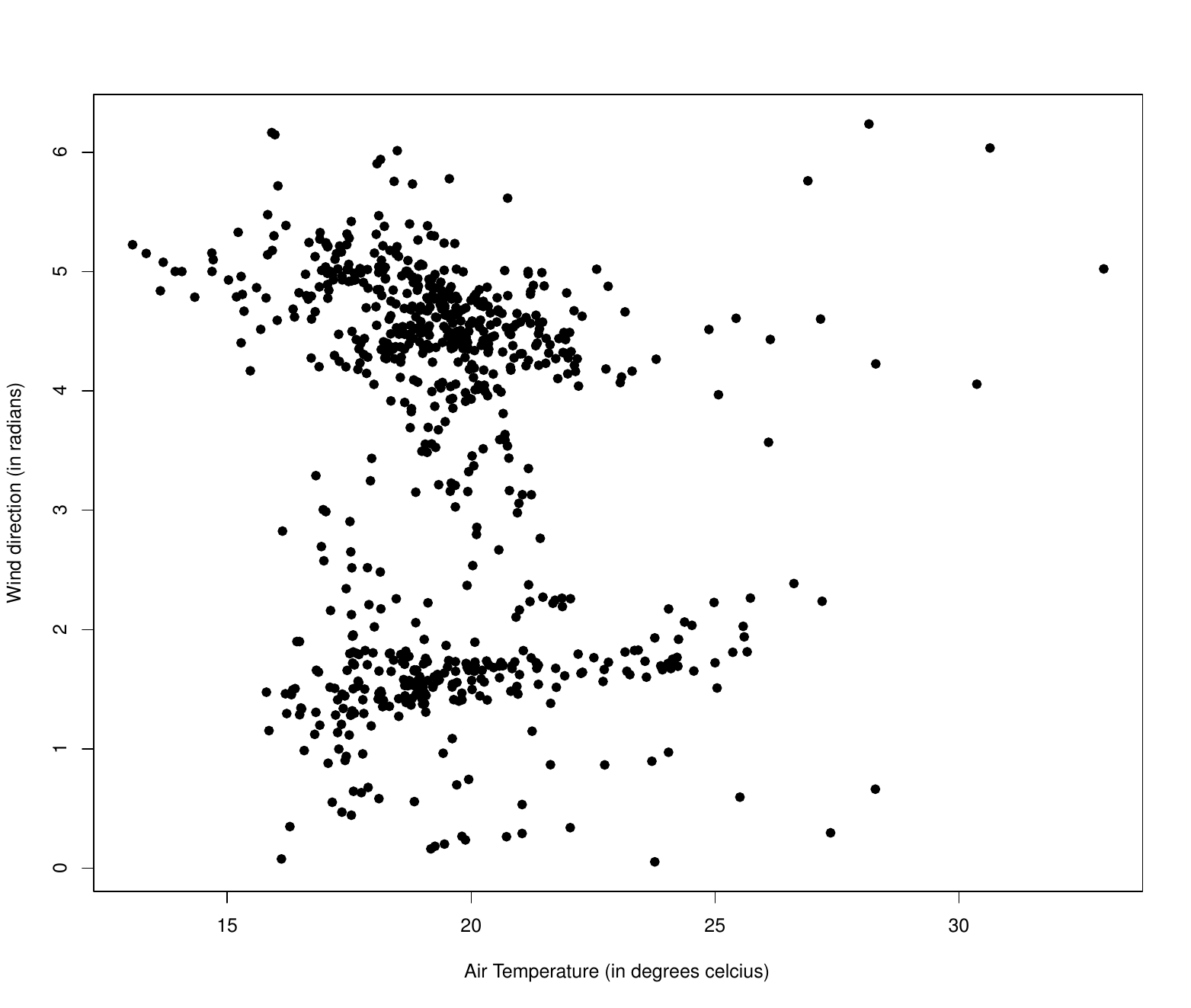}
    \caption{}
    \label{fig:scatter_windDA}
    \end{subfigure}
    \hfill
    \begin{subfigure}[b]{0.45\textwidth}
    \includegraphics[width=\textwidth]{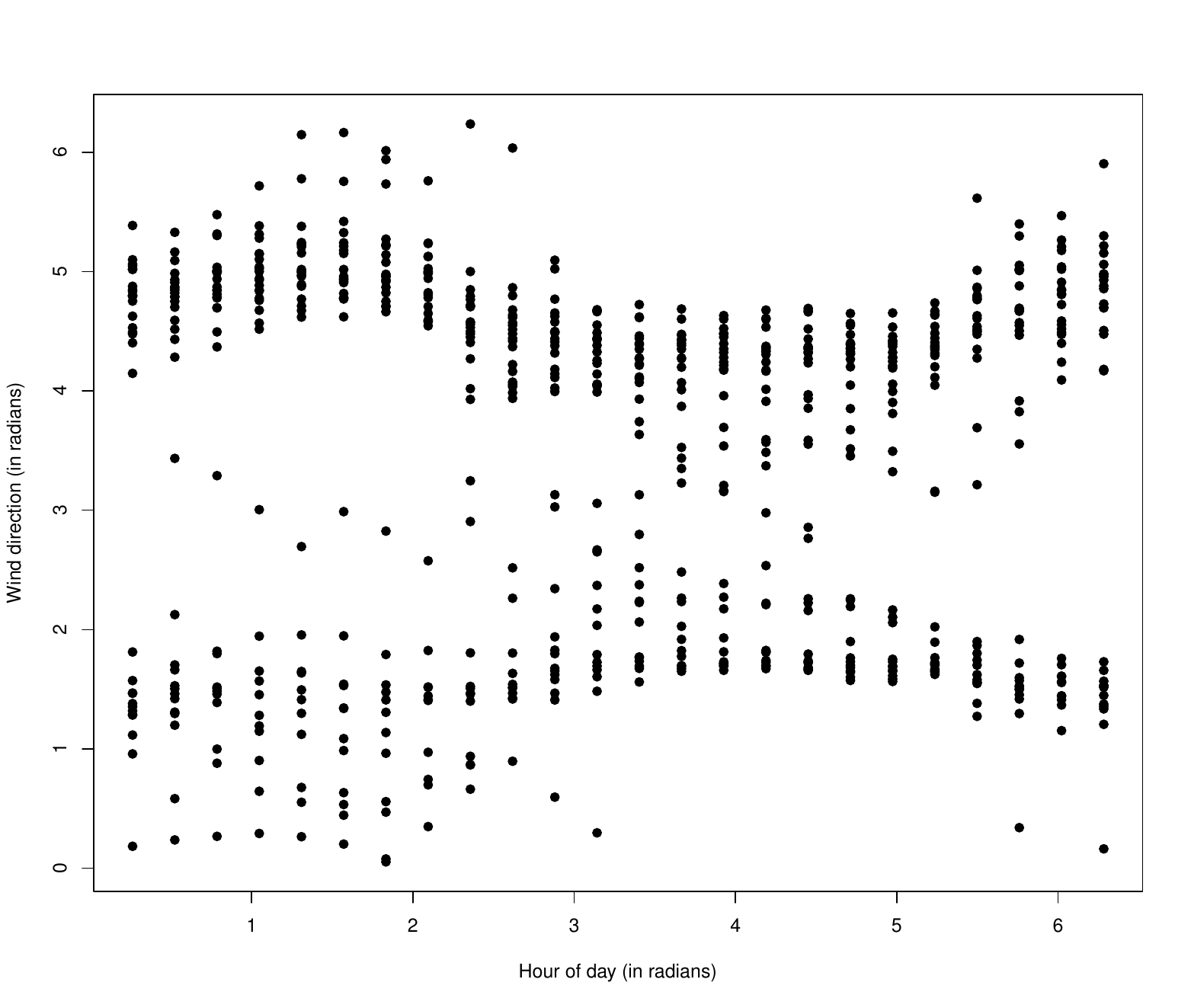}
    \caption{}
    \label{fig:scatter_windDH}
    \end{subfigure}
    \caption{(a) A circular plot of the 744 hourly average wind directions, measured in radians. (b) A scatter plot of the average wind directions and average wind speeds, in metres/second. (c) A scatter plot of the average wind directions and average air temperature, in degrees Celsius. (d) A scatter plot of the average wind directions and hour of the day, in radians. The data were recorded at a wind farm in the Eastern Cape province of South Africa.}
\end{figure}
The motivation for this paper is a study of factors that influence wind direction. The data consist of 4464 observations recorded every 10 minutes from 1 January 2019 to 31 January 2019 at a wind farm in the Eastern Cape province of South Africa. During this period, the following variables were measured: average wind direction (in radians), average wind speed (in metres/second), average air temperature (in \textdegree C), including the date and time of the measurement. To consider time on a 24-hour scale as a circular variable in this study, we decided to use hourly data and to achieve this, the data were averaged to obtain hourly data. The circular mean direction was used to calculate the hourly average wind direction and the usual arithmetic mean was used to calculate the hourly average wind speed and air temperature (\cite{jammalamadaka2006}). The resulting hourly dataset has 744 observations. To evaluate the influence of the time of day on the average wind direction, we took the hour of the day and converted it into a circular variable by multiplying it by $(2\pi/24)$. As a result, the data contain two linear covariates, wind speed and air temperature, and one circular covariate, the hour of the day.\\
Figure \ref{fig:circ_wind} shows a circular plot of the average wind directions (in radians). The observed average wind directions are clearly bimodal with one mode having a mean (wind) direction of $90$ degrees (or $\pi/2$ in radians) and another one having a mean (wind) direction of about $270$ degrees (or $3\pi/2$ in radians). Further evidence of this bimodality is apparent in the scatter plots of wind direction versus wind speed, air temperature and hour of the day shown in Figures \ref{fig:scatter_windDS}-\ref{fig:scatter_windDH}, respectively. Thus, a circular regression model, as mentioned above (see section \ref{circ_linear} for more details), is not appropriate for this data since it assumes that the circular response is unimodal.\\
The rest of this paper is structured as follows. In section \ref{methodology}, we define the von-Mises distribution, the circular-linear regression model and its estimation procedure followed by a definition of the finite mixture of circular-linear regressions and its estimation procedure. We conclude the section by discussing some modelling and computational details. In section \ref{simulations}, we conduct Monte Carlo simulations to demonstrate the use of the proposed model and the performance of the proposed estimation procedure. In section \ref{applications}, we demonstrate the proposed model's potential for practical utility by applying it to the wind data. Section \ref{discussion} concludes the paper and suggests direction for future research.
\section{Methodology}\label{methodology}
\subsection{von-Mises distribution}\label{sec_von_den}
Analogous to the Gaussian (normal) distribution for continuous variables measured on the real line, the von-Mises distribution is the most widely applicable probability distribution for variables measured on the circle (\cite{mardia2000}). In the case of a univariate circular random variable $\theta$, the density function of a von-Mises distribution, denoted by $\mathcal{VM} (\theta|\mu,\kappa)$, is 
\begin{eqnarray}\label{von_den}
    \mathcal{VM}(\theta|\mu,\kappa)=\frac{1}{2\pi I_0(\kappa)}\text{exp}(\kappa \text{cos}(\theta-\mu)),\hspace{1cm} 0<\theta\leq2\pi,
\end{eqnarray}
where $\mu\in (0,2\pi]$ is the mean direction and $\kappa\geq0$ is the concentration parameter. The function $I_0(\kappa)$ in (\ref{von_den}) is the zeroth-order modified Bessel function of the first kind (see \cite{mardia2000} and the references therein for more details).\\
The von-Mises distribution has some interesting properties. For $\kappa>0$, the density function \eqref{von_den} is unimodal and symmetrical about $\mu$. When $\kappa=0$, the density function (\ref{von_den}) reduces to a uniform density function on the circle. Lastly, as $\kappa \rightarrow \infty$, the density function \eqref{von_den} is approximately a Gaussian density function (\cite{bishop2006}). As such, the von-Mises distribution is usually referred to as the Gaussian distribution for continuous variables measured on the circle or simply circular Gaussian distribution (\cite{jammalamadaka2001} and \cite{bishop2006}). For more details about the von-Mises distribution, see Chapter 3 of \cite{mardia2000}.
\subsection{Circular regression model}\label{circ_linear}
Let $\theta\in (0,2\pi]$, $\mathbf{x}_1=(x_1,x_2,\dots,x_q)^\intercal \in (0,2\pi]^q$ and $\mathbf{x}_2=(x_1,x_2,\dots,x_p)^\intercal \in \mathbb{R}^p$ represent a circular response variable, a vector of circular covariates and a vector of linear covariates, respectively. There are three possible circular regression models: (1) the model with only linear covariates, henceforth referred to as the circular-linear regression model, (2) the model with only circular covariates, henceforth referred to as the circular-circular regression model and (3) the model with both circular and linear covariates, henceforth referred to as the circular-circular-linear regression model. We briefly discuss each model below.
\subsubsection{Circular-Linear regression model}
Given only $\mathbf{x}_2$, the conditional distribution of $\theta$ is assumed to be a von-Mises distribution, $\mathcal{VM}(\mu(\mathbf{x}_2),\kappa)$ and the conditional mean direction $\mu(\mathbf{x}_2)$ is given by
\begin{eqnarray}\label{RF1}
\mu(\mathbf{x}_2)=\mu+g(\mathbf{x}_2,\boldsymbol{\beta}_2),
\end{eqnarray}
where $\boldsymbol{\beta}_2=(\beta_1,\beta_2,\dots,\beta_p)^\intercal$ is a vector of regression coefficients and $g(\cdot)$ is a link function that maps the real line onto the circle $(-\pi,\pi)$.\\
\cite{gould1969} used the identity link function which led to a model in which the likelihood function has an infinite number of maxima with the same likelihood value (\cite{mardia2000}). In practice, this implies that $\boldsymbol{\beta}_2$ has no unique solution and hence unidentifiable. Thus, the model is not useful in practice.\\
The problems encountered with the Gould model can be avoided by making use of a monotone link function $g$ satisfying $g(0)=0$ (\cite{fisher1992}). A possible choice for a link function is the inverse tangent link function
\begin{eqnarray}\label{link1}
    g(\mathbf{x}_2,\boldsymbol{\beta}_2)=2\text{tan}^{-1}(\mathbf{x}^\intercal_2\boldsymbol{\beta}_2).
\end{eqnarray}
For other possible choices of $g$, see \cite{fisher1992} and \cite{mardia2000}.
\subsubsection{Circular-Circular regression model}
Given only $\mathbf{x}_1$, the conditional distribution of $\theta$ is assumed to be a von-Mises distribution, $\mathcal{VM}(\mu(\mathbf{x}_1),\kappa)$ and the conditional mean direction $\mu(\mathbf{x}_1)$ is given by
\begin{eqnarray}\label{RF2}
\mu(\mathbf{x}_1)&=&\mu+\sum_{j=1}^q (\beta_{1j}\text{cos}(x_j)+\beta_{2j}\text{sin}(x_j))\nonumber\\
    &=&\mu+\mathbf{x}^{\ast\intercal}_1\boldsymbol{\beta}^\ast_1,
\end{eqnarray}
where $\mathbf{x}^\ast_1=(\text{cos}(x_1),\text{sin}(x_1),\text{cos}(x_2),\text{sin}(x_2),\dots,\text{cos}(x_q),\text{sin}(x_q))$ and\\ $\boldsymbol{\beta}^\ast_1=(\beta_{11},\beta_{21},\beta_{12},\beta_{22},\dots,\beta_{1q},\beta_{2q})$.\\
Note that, as pointed out by \cite{kim2018}, the range of the regression function \eqref{RF2} may not be on the unit circle. Thus, we propose to model \eqref{RF2} using the inverse tangent link 
\begin{eqnarray}\label{RF3}
\mu(\mathbf{x}_1)&=&\mu+\text{2tan}^{-1}(\mathbf{x}^{\ast\intercal}_1\boldsymbol{\beta}^\ast_1).
\end{eqnarray}
Other forms of circular-circular regression models were proposed by \cite{sarma1993} and \cite{downs2002}, see \cite{kim2018} for a recent review.
\subsubsection{Circular-Circular-Linear regression model}
Given both $\mathbf{x}_1$ and $\mathbf{x}_2$, the conditional distribution of $\theta$ is assumed to be a von-Mises distribution, $\mathcal{VM}(\mu(\mathbf{x}_1,\mathbf{x}_2),\kappa)$. The conditional mean direction $\mu(\mathbf{x}_1,\mathbf{x}_2)$ is given by the combination of \eqref{RF1} and \eqref{RF3} as
\begin{eqnarray}\label{RF4}
\mu(\mathbf{x}_1,\mathbf{x}_2)=\mu+g(\mathbf{x}^{\ast\intercal}_1\boldsymbol{\beta}_1)+g(\mathbf{x}^\intercal_2\boldsymbol{\beta}_2),
\end{eqnarray}
where $g(x)=\text{2tan}^{-1}(x)$.\\

This implies that the Circular-Circular-Linear regression model includes the first two models as special cases. Thus, the rest of the paper focuses on the Circular-Circular-Linear regression model. To simplify the terminology, we henceforth refer to this model as the circular regression model. Moreover, to simplify the notation, let $\mathbf{x}=(\mathbf{x}_1,\mathbf{x}_2)^\intercal$, $\mathbf{x}^\ast=(\mathbf{x}^\ast_1,\mathbf{x}_2)^\intercal$ and $\boldsymbol{B}=(\boldsymbol{\beta}_1,\boldsymbol{\beta}_2)$.

\subsection{Maximum likelihood estimation}\label{MLE_circ_linear}
In this section, we present the estimation procedure for the circular regression model in \eqref{RF4}. The estimation procedures of the other two models in \eqref{RF1} and \eqref{RF3} can be obtained as special cases. Consider a random sample $\{(\mathbf{x}_i,\theta_i):i=1,2,\dots,n\}$ from the conditional distribution $\mathcal{VM}(\mu(\mathbf{x}),\kappa))$, where $\mu(\mathbf{x})$ is given by \eqref{RF4}. The log-likelihood function is
\begin{eqnarray}\label{llk}
\ell(\boldsymbol{\vartheta}|\boldsymbol{\theta},\mathbf{x}^\ast)=-n\text{log}(2\pi)-n\text{log}(I_0(\kappa))+\kappa\sum_{i=1}^n \text{cos}(\theta_i-\mu-g(\mathbf{x}^{\ast\intercal}_i\boldsymbol{B})),
\end{eqnarray}
where $\boldsymbol{\vartheta}=(\mu,\kappa,\boldsymbol{B})$ is the vector of all the model parameters.\\
The maximum likelihood estimation (MLE) of $\mu$ is obtained by maximizing (\ref{llk}) with respect to $\mu$ as
\begin{eqnarray}\label{der1_llk_mu}
\frac{\partial\ell(\boldsymbol{\vartheta}|\boldsymbol{\theta},\mathbf{x}^\ast)}{\partial\mu}=-\kappa\sum_{i=1}^n\text{sin}(\theta_i-\mu-g(\mathbf{x}^{\ast\intercal}_i\boldsymbol{B})).
\end{eqnarray}
Setting (\ref{der1_llk_mu}) equal to zero gives
\begin{eqnarray}\label{der2_llk_mu}
\sum_{i=1}^n\text{sin}(\theta_i-\mu-g(\mathbf{x}^{\ast\intercal}_i\boldsymbol{B}))=0.
\end{eqnarray}
Using the trigonometric identity
\begin{eqnarray}\label{trig_identity}
    \text{sin}(x-y)=\text{sin}(x)\text{cos}(y)-\text{sin}(y)\text{cos}(x),
\end{eqnarray}
we solve (\ref{der2_llk_mu}) for $\mu$ to obtain
\begin{eqnarray}\label{mle_mu}
\hat{\mu}=\text{tan}^{-1}\bigg[\frac{\sum_{i=1}^n\text{sin}(\theta_i-g(\mathbf{x}^{\ast\intercal}_i\boldsymbol{B}))}{\sum_{i=1}^n\text{cos}(\theta_i-g(\mathbf{x}^{\ast\intercal}_i\boldsymbol{B}))}\bigg].
\end{eqnarray}
The MLE of $\kappa$ is obtained by maximizing (\ref{llk}) with respect to $\kappa$ as
\begin{eqnarray}\label{der1_llk_kappa}
\frac{\partial\ell(\boldsymbol{\vartheta}|\boldsymbol{\theta},\mathbf{x}^\ast)}{\partial\kappa}=-nA(\kappa)+\sum_{i=1}^n\text{cos}(\theta_i-\mu-g(\mathbf{x}^{\ast\intercal}_i\boldsymbol{B})),
\end{eqnarray}
where $A(\kappa)=I_1(\kappa)/I_0(\kappa)$ and $I_1(\kappa)=I^{'}_0(\kappa)$.\\
Setting (\ref{der1_llk_kappa}) equal to zero and solving for $\kappa$ gives 
\begin{eqnarray}\label{kappa_mle}
    \hat{\kappa}=A^{-1}\bigg(\frac{1}{n}\sum_{i=1}^n\text{cos}(\theta_i-\hat{\mu}-g(\mathbf{x}^{\ast\intercal}_i\boldsymbol{B}))\bigg),
\end{eqnarray}
where $A^{-1}(x)$ is the inverse of the function $A(x)$.\\
The MLE of the regression coefficient $\boldsymbol{B}$ has no closed form expression. Thus, we approximate it using an ascent algorithm such as the Newton-Raphson (NR) algorithm. The NR update rule for $\boldsymbol{B}$ is
\begin{eqnarray}\label{NR_update1}
\boldsymbol{B}^{\text{new}}=\boldsymbol{B}^{\text{old}}-\mathbf{H}^{-1}(\boldsymbol{B}^{\text{old}})\boldsymbol{g}(\boldsymbol{B}^{\text{old}}),
\end{eqnarray}
where
\begin{eqnarray}
\boldsymbol{g}(\boldsymbol{B})&\equiv&\frac{\partial\ell(\boldsymbol{\vartheta}|\boldsymbol{\theta},\mathbf{x}^\ast)}{\partial\boldsymbol{B}}=\mathbf{X}^{\ast\intercal}\dot{\boldsymbol{G}}\mathbf{u},\\
\mathbf{H}(\boldsymbol{B})&\equiv&\frac{\partial^2\ell(\boldsymbol{\vartheta}|\boldsymbol{\theta},\mathbf{x}^\ast)}{\partial^2\boldsymbol{B}}=-\mathbf{X}^{\ast\intercal}\dot{\boldsymbol{G}}^2\boldsymbol{V}\mathbf{X}^\ast+\mathbf{X}^{\ast\intercal}\ddot{\boldsymbol{G}}\mathbf{U}\mathbf{X}^\ast,
\end{eqnarray}
with
\begin{eqnarray}
    \dot{\boldsymbol{G}}&=&\text{diag}\{g^{'}(\mathbf{x}^{\ast\intercal}_1\boldsymbol{B}),g^{'}(\mathbf{x}^{\ast\intercal}_2\boldsymbol{B}),\dots,g^{'}(\mathbf{x}^{\ast\intercal}_n\boldsymbol{B})\},\nonumber\\
    \ddot{\boldsymbol{G}}&=&\text{diag}\{g^{''}(\mathbf{x}^{\ast\intercal}_1\boldsymbol{B}),g^{''}(\mathbf{x}^{\ast\intercal}_2\boldsymbol{B}),\dots,g^{''}(\mathbf{x}^{\ast\intercal}_n\boldsymbol{B})\},\nonumber\\
    \mathbf{X}^\ast&=&(\mathbf{x}^{\ast\intercal}_1,\mathbf{x}^{\ast\intercal}_2,\dots,\mathbf{x}^{\ast\intercal}_n)^\intercal,\nonumber\\
    \mathbf{u}&=&(u_1,u_2,\dots,u_n),\nonumber\\
    \mathbf{U}&=&\text{diag}\{u_1,u_2,\dots,u_n\},\nonumber\\
    \mathbf{V}&=&\text{diag}\{v_1,v_2,\dots,v_n\},\nonumber\\
    u_i&=&\text{sin}(\theta_i-\hat{\mu}-g(\mathbf{x}^{\ast\intercal}_i\boldsymbol{B})),\nonumber\\
    v_i&=&\text{cos}(\theta_i-\hat{\mu}-g(\mathbf{x}^{\ast\intercal}_i\boldsymbol{B}))\hspace{0.2cm} \text{for $i=1,2,\dots,n$}.\nonumber
\end{eqnarray}
Note that $g^{'}(x)$ and $g^{''}(x)$ are the first and second derivatives of the function $g(x)$, respectively.\\
The MLE of $\boldsymbol{B}$, denoted $\hat{\boldsymbol{B}}$, is the updated value of $\boldsymbol{B}^{\text{old}}$ at convergence of the NR algorithm. Notice that, in order to obtain both $\hat{\mu}$ and $\hat{\kappa}$, we need to know $\boldsymbol{B}$. Following the suggestion in \cite{fisher1992}, we obtain $\hat{\mu}$, $\hat{\kappa}$ and $\hat{\boldsymbol{B}}$ using an iterative scheme. We begin with an initial value of $\boldsymbol{B}$, denoted $\boldsymbol{B}^{(0)}$, and use it to obtain $\hat{\mu}$ and hence $\hat{\kappa}$. Using $\hat{\mu}$, then $\boldsymbol{B}^{(0)}$ is updated applying \eqref{NR_update1} to obtain $\hat{\boldsymbol{B}}$. We use $\hat{\boldsymbol{B}}$ to update $\hat{\mu}$ and $\hat{\kappa}$ and repeat this cycle until convergence. The above estimation procedure is summarised in Algorithm \ref{alg:Algorithm1}.
\begin{algorithm}
\caption{Fitting a circular-linear regression model}
\label{alg:Algorithm1}
\begin{algorithmic}[1]
    \STATE Let $\boldsymbol{B}^{(0)}$ be an appropriately chosen the initial value for $\boldsymbol{B}$.
    \STATE Given $\hat{\boldsymbol{B}}^{(r-1)}$, at the $r^{th}$ iteration, estimate $\mu$ and $\kappa$ using \eqref{mle_mu} and \eqref{kappa_mle}.
    \STATE Given $\hat{\mu}^{(r)}$ from Step 2, update $\boldsymbol{B}^{(r-1)}$ using \eqref{NR_update1}.
    \STATE Repeat Step 2 and Step 3 until convergence.
\end{algorithmic}
\end{algorithm}
\subsection{Mixtures of circular regressions}
Suppose that the random sample $\{\theta_1,\theta_2,\dots,\theta_n\}$ has been drawn from a population that consists of $K$ sub-populations or components, $1,2,\dots,K$, each comprising an unknown proportion $\pi_1,\pi_2,\dots,\pi_K$ of data points generated from a $\mathcal{VM}(\mu_k,\kappa_k)$ sub-population, for $k=1,2,\dots,K$. Therefore, $\theta$ is said to be generated from a mixture of von-Mises distributions
\begin{eqnarray}\label{mix_von_mises}
\mathcal{MVM}(\theta)=\sum_{k=1}^K\pi_k \mathcal{VM}(\mu_k,\kappa_k),
\end{eqnarray}
where $\pi_k>0$, for $k=1,2,\dots,K$, and satisfying $\sum_{k=1}^K\pi_k=1$, are the mixing proportions. The parameters $\mu_k$ and $\kappa_k$ are the mean direction and concentration, respectively, for the $k^{th}$ component.\\
Suppose that, in addition to $\theta_i$, we observe a vector of covariates $\mathbf{x}_i=(\mathbf{x}_{1i},\mathbf{x}_{2i})\in (0,2\pi]^q\times\mathbb{R}^p$, for $i=1,2,\dots,n$, where $\mathbf{x}_1$ represents a vector of $q$ circular covariates and $\mathbf{x}_2$ represents a vector of $p$ linear covariates. To model the dependence of $\theta$ on $\mathbf{x}$, we propose a mixture of circular regressions in which the conditional distribution of the circular response variable $\theta$, conditional on $\mathbf{X}=\mathbf{x}$, is given by the mixture of von-Mises distributions 
\begin{eqnarray}\label{mix_circ}
\mathcal{MVM}(\theta|\mathbf{X}=\mathbf{x})=\sum_{k=1}^K\pi_k \mathcal{VM}(\mu_k+g(\mathbf{x}^{\ast\intercal}\boldsymbol{B}_k)),\kappa_k),
\end{eqnarray}
where $\mathbf{x}^\ast=(\mathbf{x}^\ast_1,\mathbf{x}_2)^\intercal$, with $\mathbf{x}^\ast_1=(\text{cos}(x_1),\text{sin}(x_1),\text{cos}(x_2),\text{sin}(x_2),\dots,\text{cos}(x_q),\text{sin}(x_q))$, and $\boldsymbol{B}_k$ is the vector of regression coefficients for the $k^{th}$ component.\\
In this paper, we assume that $K$ is fixed. Thus, model (\ref{mix_circ}) is a finite mixture of circular regressions. However, in most practical applications $K$ is unknown and must be inferred from the data. In section \ref{number_components}, we propose a data-driven approach to choose $K$.
\subsection{MLE via the Expectation-Maximization algorithm}
Consider a random sample $\{(\mathbf{x}_i,\theta_i):i=1,2,\dots,n\}$ from model (\ref{mix_circ}) and define $\mathbf{x}^\ast_i$, for $i=1,2,\dots,n$. The log-likelihood function is given by
\begin{eqnarray}\label{llk2}
\ell(\boldsymbol{\vartheta}|\boldsymbol{\theta},\mathbf{x}^\ast)=\sum_{i=1}^n\text{log}\sum_{k=1}^K\pi_k\mathcal{VM}(\mu_k+g(\mathbf{x}^{\ast\intercal}_i\boldsymbol{B}_k),\kappa_k),
\end{eqnarray}
where $\boldsymbol{\vartheta}=((\pi_1,\mu_1,\kappa_1,\boldsymbol{B}_1),(\pi_2,\mu_2,\kappa_2,\boldsymbol{B}_2),\dots,(\pi_K,\mu_K,\kappa_K,\boldsymbol{B}_K))$ is a vector of all the model parameters.\\
It is well-known that a log-likelihood function of the form (\ref{llk2}) cannot be maximized analytically to obtain the MLE of $\boldsymbol{\vartheta}$. Thus, we maximize \eqref{llk2} numerically using the Expectation-Maximization (EM) algorithm (\cite{DLR1977}). Towards that end, for each $(\mathbf{x}^\ast_i,\theta_i)$, for $i=1,2,\dots,n$, we define the latent variable $\mathbf{z}_i=(z_{i1},z_{i2},\dots,z_{iK})$ such that
\begin{eqnarray}
z_{ik}=\begin{cases}
    0\hspace{1cm} \text{if $(\mathbf{x}^\ast_i,\theta_i)$ is not generated from the $k^{th}$ component},\nonumber\\
    1\hspace{1cm} \text{if $(\mathbf{x}^\ast_i,\theta_i)$ is generated from the $k^{th}$ component}.
\end{cases}
\end{eqnarray}
Given the complete data $\{(\mathbf{x}^\ast_i,\theta_i,\mathbf{z}_i):i=1,2,\dots,n\}$, the log-likelihood is obtained as
\begin{eqnarray}
\ell^c(\boldsymbol{\vartheta}|\boldsymbol{\theta},\mathbf{x}^\ast)=\sum_{i=1}^n\sum_{k=1}^K z_{ik}\bigg[\text{log}\pi_k+\text{log}\mathcal{VM}(\theta_i|\mu_k+g(\mathbf{x}^{\ast\intercal}_i\boldsymbol{B}_k),\kappa_k)\bigg].
\end{eqnarray}
Given a suitable initial value for $\boldsymbol{\vartheta}$, denoted $\boldsymbol{\vartheta}^{(0)}$, the EM algorithm iterates between two steps, the Expectation (E-) step and the Maximization (M-) step. In the E-step, at the $r^{th}$ iteration, the algorithm obtains the expected value of $\ell^c(\boldsymbol{\vartheta}|\boldsymbol{\theta},\mathbf{x}^\ast)$, based on the conditional distribution of $\mathbf{z}$ given $\mathbf{x}^\ast$, $\boldsymbol{\theta}$ and $\boldsymbol{\vartheta}^{(r-1)}$, as
\begin{eqnarray}\label{exp_llk}
    Q(\boldsymbol{\vartheta}|\boldsymbol{\vartheta}^{(r-1)})=\sum_{i=1}^n\sum_{k=1}^K \mathbb{E}[z_{ik}|\mathbf{x}_i,\theta_i,\boldsymbol{\vartheta}^{(r-1)}]\bigg[\text{log}\pi_k+\text{log}\mathcal{VM}(\theta_i|\mu_k+g(\mathbf{x}^{\ast\intercal}_i\boldsymbol{B}_k),\kappa_k)\bigg],
\end{eqnarray}
which reduces to calculating $\mathbb{E}[z_{ik}|\mathbf{x}^\ast_i,\theta_i,\boldsymbol{\vartheta}^{(r-1)}]$, for $i=1,2,\dots,n$, as
\begin{eqnarray}\label{resp}
    \gamma^{(r)}_{ik}=\frac{\pi_{k}^{(r-1)}\mathcal{VM}(\theta_i|\mu^{(r-1)}_k+g(\mathbf{x}^{\ast\intercal}_i\boldsymbol{B}^{(r-1)}_k),\kappa^{(r-1)}_k)}{\sum_{j=1}^K\pi_{j}^{(r-1)}\mathcal{VM}(\theta_i|\mu^{(r-1)}_j+g(\mathbf{x}^{\ast\intercal}_i\boldsymbol{B}^{(r-1)}_j),\kappa^{(r-1)}_j)}.
\end{eqnarray}
In the M-step, we maximize $Q(\boldsymbol{\vartheta}|\boldsymbol{\vartheta}^{(r-1)})$ to update $\boldsymbol{\vartheta}^{(r-1)}$, after setting\\
$\mathbb{E}[z_{ik}|\mathbf{x}^\ast_i,\theta_i,\boldsymbol{\vartheta}^{(r-1)}]=\gamma^{(r)}_{ik}$, for $i=1,2,\dots,n$.\\
Maximizing $Q(\boldsymbol{\vartheta}|\boldsymbol{\vartheta}^{(r-1)})$ with respect to $\mu_k$, for $k=1,2,\dots,K$, gives
\begin{eqnarray}\label{der1_mu_mix}
    \frac{\partial Q(\boldsymbol{\vartheta}|\boldsymbol{\vartheta}^{(r-1)})}{\partial \mu_k}=-\kappa_k\sum_{i=1}^n\gamma^{(r)}_{ik}\text{sin}(\theta_i-\mu_k-g(\mathbf{x}^{\ast\intercal}_i\boldsymbol{B}_k)).
\end{eqnarray}
Setting \eqref{der1_mu_mix} equal to zero and solving for $\mu_k$, using the trigonometric identity (\ref{trig_identity}), gives the update for $\mu^{(r-1)}_{k}$, for $k=1,2,\dots,K$, as
\begin{eqnarray}\label{mix_mle_mu}
\mu^{(r)}_k=\text{tan}^{-1}\bigg[\frac{\sum_{i=1}^n\gamma^{(r)}_{ik}\text{sin}(\theta_i-g(\mathbf{x}^{\ast\intercal}_i\boldsymbol{B}^{(r-1)}_k))}{\sum_{i=1}^n\gamma^{(r)}_{ik}\text{cos}(\theta_i-g(\mathbf{x}^.{\ast\intercal}_i\boldsymbol{B}^{(r-1)}_k))}\bigg].
\end{eqnarray}
Maximizing $Q(\boldsymbol{\vartheta}|\boldsymbol{\vartheta}^{(r-1)})$ with respect to $\kappa_k$, for $k=1,2,\dots,K$, gives
\begin{eqnarray}\label{der1_kappa_mix}
    \frac{\partial Q(\boldsymbol{\vartheta}|\boldsymbol{\vartheta}^{(r-1)})}{\partial \kappa_k}=-n_kA(\kappa_k)+\sum_{i=1}^n\gamma^{(r)}_{ik}\text{cos}(\theta_i-\mu_k-g(\mathbf{x}^\intercal_i\boldsymbol{\beta}_k)),
\end{eqnarray}
where $n_k=\gamma^{(r)}_{ik}$, for $k=1,2,\dots,K$.\\
Setting \eqref{der1_kappa_mix} equal to zero and solving for $\kappa_k$ gives the update equation for $\kappa^{(r-1)}_k$, for $k=1,2,\dots,K$, as
\begin{eqnarray}\label{mle_kappa_mix}
    \kappa^{(r)}_k=A^{-1}\bigg(\frac{1}{n_k}\sum_{i=1}^n\gamma^{(r)}_{ik}\text{cos}(\theta_i-\mu^{(r)}_k-g(\mathbf{x}^{\ast\intercal}_i\boldsymbol{B}^{(r-1)}_k))\bigg).
\end{eqnarray}
To update the regression coefficients $\boldsymbol{B}^{(r-1)}_k$, for $k=1,2,\dots,K$, we iteratively maximize $Q(\boldsymbol{\vartheta}|\boldsymbol{\vartheta}^{(r-1)})$, with respect to $\boldsymbol{B}_k$, for $k=1,2,\dots,K$, using the NR algorithm.\\
The NR update rule for $\boldsymbol{B}_k$, for $k=1,2,\dots,K$, is 
\begin{eqnarray}\label{NR_update2}
    \boldsymbol{B}_k^{\text{new}}=\boldsymbol{B}_k^{\text{old}}-\mathbf{H}^{-1}(\boldsymbol{B}_k^{\text{old}})\boldsymbol{g}(\boldsymbol{B}_k^{\text{old}}),
\end{eqnarray}
where 
\begin{eqnarray}
\boldsymbol{g}(\boldsymbol{B}_k)&\equiv&\frac{\partial Q(\boldsymbol{\vartheta}|\boldsymbol{\vartheta}^{(r-1)})}{\partial \boldsymbol{B}_k}=\mathbf{X}^{\ast\intercal}\dot{\boldsymbol{G}}_k\mathbf{W}^{(r)}_k\mathbf{u}_k,\\
\mathbf{H}(\boldsymbol{B}_k)&\equiv&\frac{\partial^2 Q(\boldsymbol{\vartheta}|\boldsymbol{\vartheta}^{(r-1)})}{\partial^2 \boldsymbol{B}_k}=-\mathbf{X}^{\ast\intercal}\dot{\boldsymbol{G}}_k^2\mathbf{W}^{(r)}_k\boldsymbol{V}_k\mathbf{X}^\ast+\mathbf{X}^{\ast\intercal}\ddot{\boldsymbol{G}}_k\mathbf{W}^{(r)}_k\mathbf{U}_k\mathbf{X}^\ast,
\end{eqnarray}
with $\boldsymbol{W}^{(r)}_k=\text{diag}\{\gamma^{(r)}_{1k},\gamma^{(r)}_{2k},\dots,\gamma^{(r)}_{nk}\}$, for $k=1,2,\dots,K$.\\
For each $k=1,2,\dots,K$, $\dot{\boldsymbol{G}}_k$, $\mathbf{u}_k$, $\boldsymbol{V}_k$, $\ddot{\boldsymbol{G}}_k$ and $\mathbf{U}_k$ are defined in a similar manner as in Section \ref{MLE_circ_linear}. To update $\boldsymbol{B}^{(r-1)}_k$, we use $\boldsymbol{B}^{(r)}_k$, which is the updated value of $\boldsymbol{B}^{\text{old}}_k$, for $k=1,2,\dots,K$, at convergence of the NR algorithm. We repeat the above E- and M-steps until convergence. The above estimation procedure is summarised in Algorithm \ref{alg:Algorithm2}.
\begin{algorithm}
\caption{Fitting a mixture of circular-linear regressions model (\ref{mix_circ})}
\label{alg:Algorithm2}
\begin{algorithmic}[1]
    \STATE Let $\mu^{(0)}_k$, $\kappa^{(0)}_k$ and $\boldsymbol{B}_k^{(0)}$ be appropriately chosen the initial values for $\mu_k$, $\kappa_k$ and $\boldsymbol{B}_k$, respectively, for $k=1,2,\dots,K$.
    \STATE\textbf{E-step:} At the $r^{th}$ iteration, calculate the posterior probabilities $\gamma^{(r)}_{ik}$ using (\ref{resp}), for $i=1,2,\dots,n$ and $k=1,2,\dots,K$.
    \STATE\textbf{M-step:} At the $r^{th}$ iteration, update $\mu^{(r-1)}_k$, $\kappa^{(r-1)}_k$ and thereafter $\boldsymbol{B}^{(r-1)}_k$ using (\ref{mix_mle_mu}), (\ref{mle_kappa_mix}) and (\ref{NR_update2}), respectively, for $k=1,2,\dots,K$.
    \STATE Repeat the \textbf{E-step} and \textbf{M-step} until convergence.
\end{algorithmic}
\end{algorithm}
\subsection{Modelling and computational details}
\subsubsection{Model-based clustering rule}\label{MBC_rule}
For each observation $(\mathbf{x}_i,\theta_i)$, for $i=1,2,\dots,n$, generated from model (\ref{mix_circ}), let $\mathbf{z}_i=(z_{i1},z_{i2},\dots,z_{iK})$ be the cluster membership indicator vector, where $z_{ik}=1$ if $(\mathbf{x}_i,\theta_i)$ belongs to cluster $k$ and $z_{ik}=0$ otherwise.\\
Let $\hat{\boldsymbol{\vartheta}}$ be the MLE of $\boldsymbol{\vartheta}$ obtained by maximizing the log-likelihood \eqref{llk2}. We can estimate each $\mathbf{z}_i$, for $i=1,2,\dots,n$, using maximum a posteriori (MAP) as
\begin{eqnarray}
\hat{z}_{ik}=\begin{cases}
    1\hspace{1cm} \text{if $\max_j\{\gamma_{ij}\}=\gamma_{ik}$}\nonumber\\
    0\hspace{1cm} \text{otherwise}
\end{cases}
\end{eqnarray}
where
\begin{eqnarray}
    \gamma_{ik}=\frac{\hat{\pi}_{k}\mathcal{VM}(\theta_i|\hat{\mu}_k+g(\mathbf{x}^{\ast\intercal}_i\boldsymbol{\hat{B}}_k),\hat{\kappa}_k)}{\sum_{j=1}^K\hat{\pi}_j\mathcal{VM}(\theta_i|\hat{\mu}_j+g(\mathbf{x}^{\ast\intercal}_i\boldsymbol{\hat{B}}_j),\hat{\kappa}_j)},
\end{eqnarray}
is the posterior probability that $(\mathbf{x}_i,\theta_i)$ belongs to the $k^{th}$ component.
\subsubsection{Choosing the number of components}\label{number_components}
In practice, prior to estimating the model in \eqref{mix_circ}, we need to choose the number of components $K$. This is a difficult task for most practical problems and failure to choose a reasonable value of $K$ may lead to a poor fitting model, characterised by an underfitted value of $K$, or complications with identifiability as a result of an overfitted value of $K$ (\cite{fruhwirth2019}; Ch. 7). For most real-world problems, $K$ can be chosen \emph{a priori}, that is before obtaining the observed data. For this reason, data-driven approaches, such as the information criteria, are employed to select $K$. In this paper, we propose to make use of the Bayesian information criterion (BIC) (\cite{schwarz1978})
\begin{eqnarray}\label{BIC}
    \text{BIC}(K)=-2\ell(\hat{\boldsymbol{\vartheta}})+\text{log}(n)\times \text{df}(K),
\end{eqnarray}
where $\hat{\boldsymbol{\vartheta}}$ is the MLE of $\boldsymbol{\vartheta}$ and $\text{df}(K)$ is the number of estimated parameters for the fitted $K-$component model.\\
The best model is the one that minimises the BIC ($K$) (\ref{BIC}), for $K\in \{1,2,\dots,K_{max}\}$, where $K_{max}$ is the largest value of $K$ considered. Note that, for $K=1$, the mixture of circular-linear regressions (\ref{mix_circ}) corresponds to the circular-linear regression model.
\subsubsection{Computing the mean direction}
Let $S=\sum_{i=1}^n \text{sin}(\theta_i-g(\mathbf{x}^{\ast\intercal}_i\boldsymbol{B}))$ and $C=\sum_{i=1}^n \text{cos}(\theta_i-g(\mathbf{x}^{\ast\intercal}_i\boldsymbol{B}))$ be the numerator and denominator, respectively, of the term in the inverse tangent of (\ref{mle_mu}). According to \cite{fisher1995} and \cite{jammalamadaka2001}, the correct estimate of $\mu$ on $[0,2\pi)$ is one which takes into account the signs of $S$ and $C$. Thus, $\hat{\mu}$ in \eqref{mle_mu} is defined as
\begin{eqnarray}\label{mle_mu2}
    \hat{\mu}^\ast=\begin{cases}
        \text{tan}^{-1}(S/C)\hspace{1.2cm} \text{if $S>0$ and $C>0$},\\
        \text{tan}^{-1}(S/C)+\pi\hspace{0.6cm} \text{if $C<0$},\\
        \text{tan}^{-1}(S/C)+2\pi\hspace{0.5cm} \text{if $S<0$ and $C>0$}.\\
    \end{cases}
\end{eqnarray}
The above also applies to $\mu^{(r)}_k$ in \eqref{mix_mle_mu}, for $k=1,2,\dots,K$, which is defined in a similar manner as in \eqref{mle_mu2} as $\mu^{\ast(r)}_k$, for $k=1,2,\dots,K$, where $S=S^{(r)}_k=\sum_{i=1}^n\gamma^{(r)}_{ik} \text{sin}(\theta_i-g(\mathbf{x}^{\ast\intercal}_i\boldsymbol{B}^{(r-1)}_k))$ and $C=C_k^{(r)}=\sum_{i=1}^n\gamma^{(r)}_{ik} \text{cos}(\theta_i-g(\mathbf{x}^{\ast\intercal}_i\boldsymbol{B}^{(r-1)}_k))$, for $k=1,2,\dots,K$.
\subsubsection{Initialization strategy}
The dependence of the EM algorithm on its initial values, for mixture modelling in particular, is well-documented in the literature (\cite{shireman2017}) and the problem has received considerable attention which resulted in a number of proposals (\cite{biernacki2003} and Chapter 2 of \cite{fruhwirth2019}). In this paper, we make use of a random initialization strategy. We initialize the algorithm from $R$ different, randomly chosen, parameter values and choose the solution that results in the largest log-likelihood value. We set the value of $R=10$ and $R=50$ in our simulations and real data analysis, respectively. Based on our simulation results (see section \ref{simulations}), this strategy works well. 
\section{Simulation studies}\label{simulations}
In this section, we demonstrate the performance of the proposed methods through extensive Monte Carlo simulations. We consider various scenarios such as increasing the number of components and also varying degrees of overlap in the components. For each scenario, we generate $1000$ samples of sizes $n=500$, $1000$ and $2000$. For illustrative purposes, all our scenarios make make use of one linear covariate $\mathbf{x}_2=x$, generated from a uniform distribution on the interval $(-0.5,0.5)$, and one circular covariate $\mathbf{x}_1=(\text{sin}(x),\text{cos}(x))$, generated from a uniform distribution on the interval $(\pi/3,8\pi/3)$. Hence, $\mathbf{x}^\ast=(\mathbf{x}_1,\mathbf{x}_2)^\intercal$ in model (\ref{mix_circ}), the latter henceforth referred to as the MixCircReg model.
\subsection{Estimation performance}\label{est_perf}
We make use of the root mean squared error (RMSE) to evaluate the performance of the estimated model parameters.
\subsubsection{Scenario 1: $K=2$ non-overlapping components}
\begin{figure}[!t]
    \centering
    \begin{subfigure}{0.48\textwidth}
        \centering
        \includegraphics[width=\textwidth]{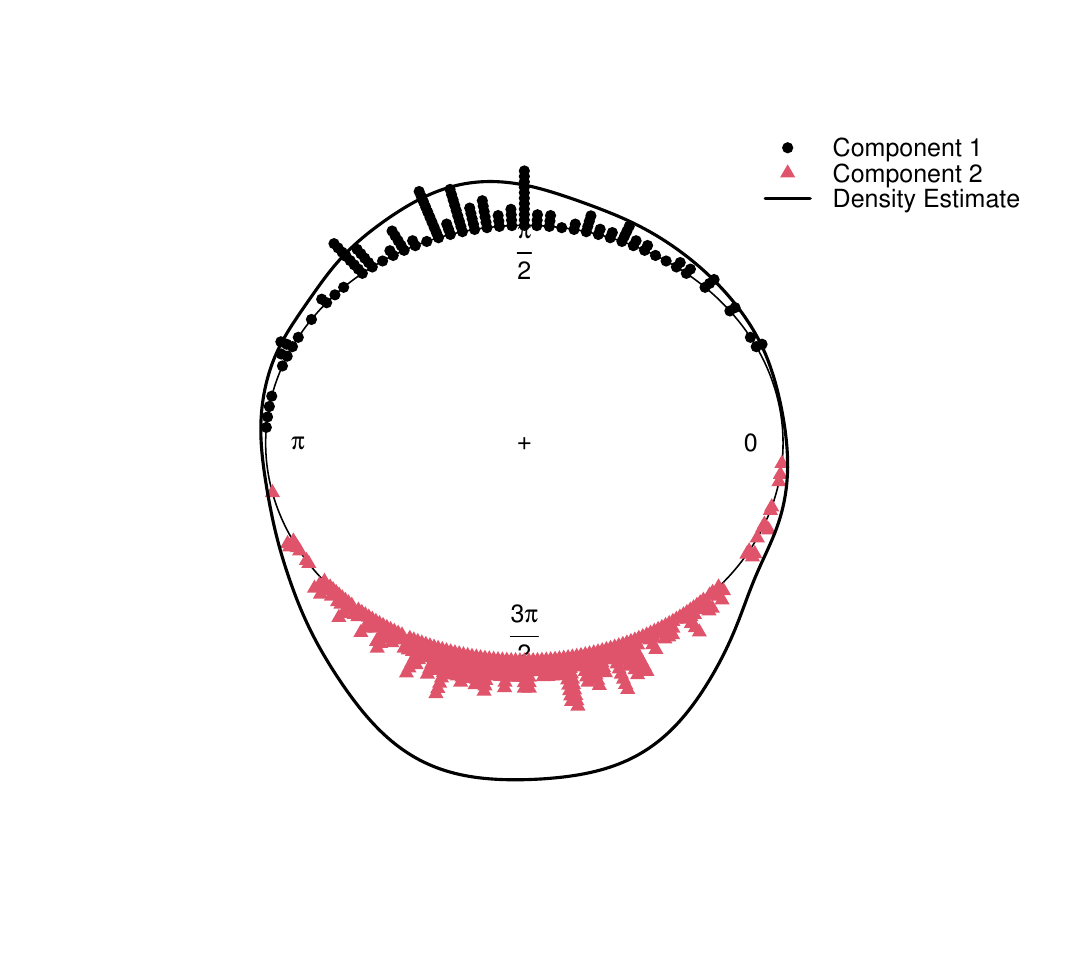}
        \caption{Non-overlapping components}
        \label{fig:plot_1_K=2}
    \end{subfigure}
    \hfill
    \begin{subfigure}{0.48\textwidth}
        \centering
        \includegraphics[width=\textwidth]{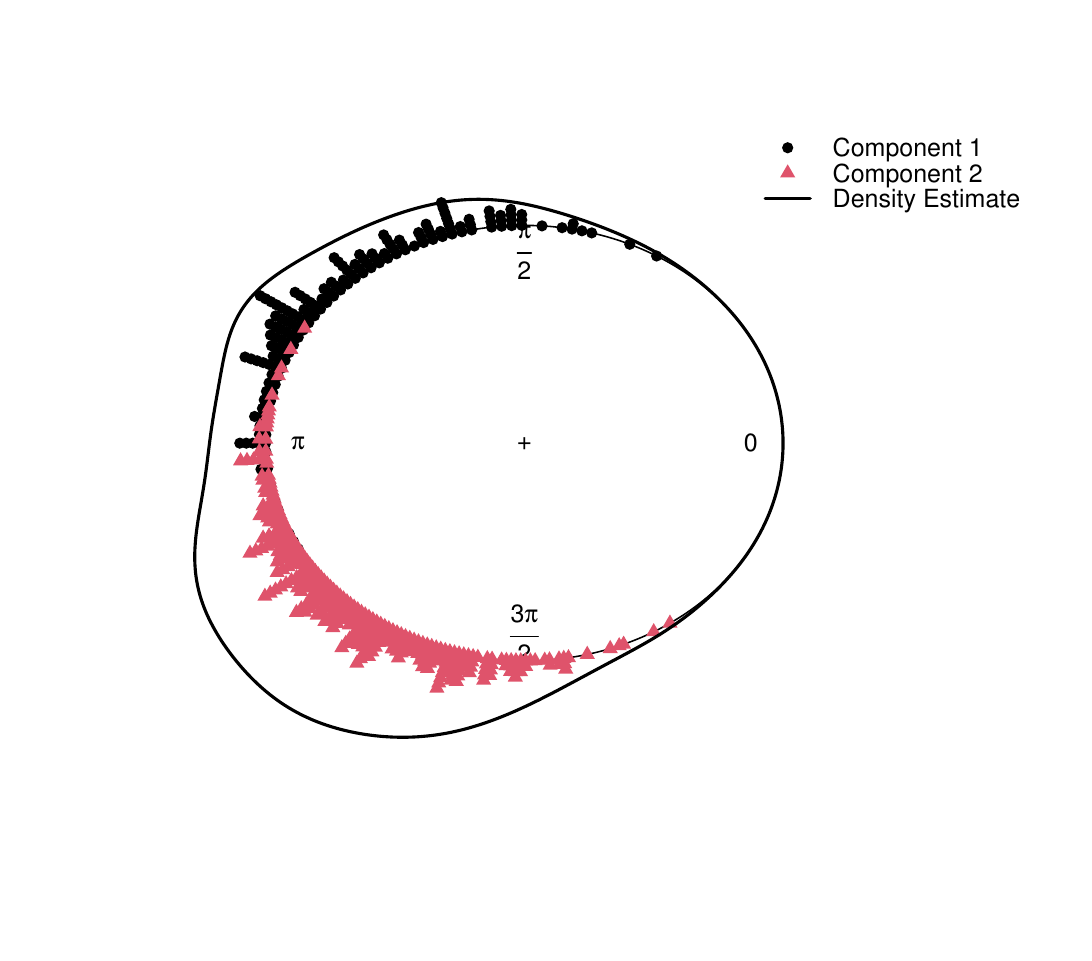}
        \caption{Overlapping components}
        \label{fig:plot_2_K=2}
    \end{subfigure}
    \caption{Circular histograms of a typical sample of size $n=500$ generated using \eqref{sim1_data} for the (a) non-overlapping and (b) overlapping case.}
    \label{fig:1}
\end{figure}

For our first example, we demonstrate the performance of the proposed methods in detecting two non-overlapping components. The data is generated as follows
\begin{eqnarray}\label{sim1_data}
    \theta_i&=\begin{cases}
        \mathcal{VM}(\mu_1+g(\mathbf{x}_i^{*\top}\boldsymbol{B}_1),\kappa_1)\hspace{0.2cm} \text{with probability}\hspace{0.2cm} \pi\\
        \mathcal{VM}(\mu_2+g(\mathbf{x}_i^{*\top}\boldsymbol{B}_2),\kappa_2)\hspace{0.2cm} \text{with probability}\hspace{0.2cm} 1-\pi
    \end{cases}\\
    &\text{for $i=1,2,\dots,n$}\nonumber
\end{eqnarray}
where $\mu_1=1.8850$, $\mu_2=4.7124$, $\kappa_1=4$, $\kappa_2=6$, $\pi=0.3$, $\boldsymbol{B}_1=(\beta_{11},\beta_{12},\beta_{13})^\top$, with $\beta_{11}=0.2$, $\beta_{12}=0.1$ and $\beta_{13}=0.3$, and $\boldsymbol{B}_2=(\beta_{21},\beta_{22},\beta_{23})$, with $\beta_{21}=0.1$, $\beta_{22}=0.2$ and $\beta_{23}=0.2$. Figure \ref{fig:plot_1_K=2} shows a circular histogram (with a kernel density estimate overlaid on the plot) of an example of data generated using \eqref{sim1_data} for $n=500$. It is clear from Figure \ref{fig:plot_1_K=2} that the two components do not overlap. Table \ref{tab:ex1} gives the results obtained from fitting the MixCircReg model to all the $1000$ generated samples of sizes $n=500$, $1000$ and $2000$. It can be seen that the estimation error is decreasing as the sample size increases which illustrates the convergence property of the proposed maximum likelihood estimators.

\begin{table}[!ht]
\small
    \centering
    \caption{RMSE of the parameter estimates for Scenario 1 and Scenario 2 over the 1000 replications}
    \begin{tabular}{||c|ccc|ccc||}
    \hline
    Parameter&\multicolumn{3}{c}{Scenario 1}&\multicolumn{3}{|c|}{Scenario 2}\\
    \hline
    &$n=500$&$n=1000$&$n=2000$&$n=500$&$n=1000$&$n=2000$\\
    \hline
    $\pi$&0.0206&0.0148&0.0105&0.0233&0.0184&0.0147\\
    \hline
    $\mu_1$&0.0617&0.0461&0.0342&0.0803&0.0581&0.0393\\
    \hline
    $\mu_2$&0.0331&0.0242&0.0171&0.0403&0.0298&0.0223\\
    \hline
    $\kappa_1$&0.5018&0.3410&0.2282&0.6699&0.4643&0.3764\\
    \hline
    $\kappa_2$&0.4622&0.3839&0.2527&0.5289&0.4344&0.2965\\
    \hline
    $\beta_{11}$&0.0507&0.0383&0.0282&0.0506&0.0417&0.0271\\
     \hline
    $\beta_{12}$&0.0541&0.0492&0.0456&0.0557&0.0482&0.0502\\
    \hline
    $\beta_{13}$&0.1344&0.1267&0.1221&0.1403&0.1232&0.1335\\
        \hline
    $\beta_{21}$&0.0541&0.0192&0.0141&0.0287&0.0246&0.0165\\
         \hline
    $\beta_{22}$&0.0489&0.0427&0.0336&0.0495&0.0445&0.0362\\
        \hline
    $\beta_{23}$&0.1267&0.1146&0.0902&0.1282&0.1231&0.0964\\
         \hline
    \end{tabular}
    \label{tab:ex1}
\end{table}

\subsubsection{Scenario 2: $K=2$ overlapping components}
We now demonstrate the performance of the proposed methods in detecting two overlapping components. The data is generated using \eqref{sim1_data} with $\mu_1=2.5133$ and $\mu_2=4.0841$. The rest of the parameter values are the same as in Scenario 1. Figure \ref{fig:plot_2_K=2} shows a circular histogram (with a kernel density estimate overlaid on the plot) of an example of data generated using \eqref{sim1_data} for $n=500$. It is clear from Figure \eqref{fig:plot_2_K=2} that the two components are overlapping.\\
Table \ref{tab:ex2} gives the results obtained from fitting the MixCircReg model to all the $1000$ generated samples of sizes $n=500$, $1000$ and $2000$. The results can be interpreted in a similar as the ones obtained for scenario 1. However, compared to scenario 1, it can be seen that the convergence rate is slower when the components are overlapping. 
\subsubsection{Scenario 3: $K=3$ non-overlapping components}
\begin{figure}[ht]
    \centering
    \begin{subfigure}{0.48\textwidth}
        \centering
        \includegraphics[width=\textwidth]{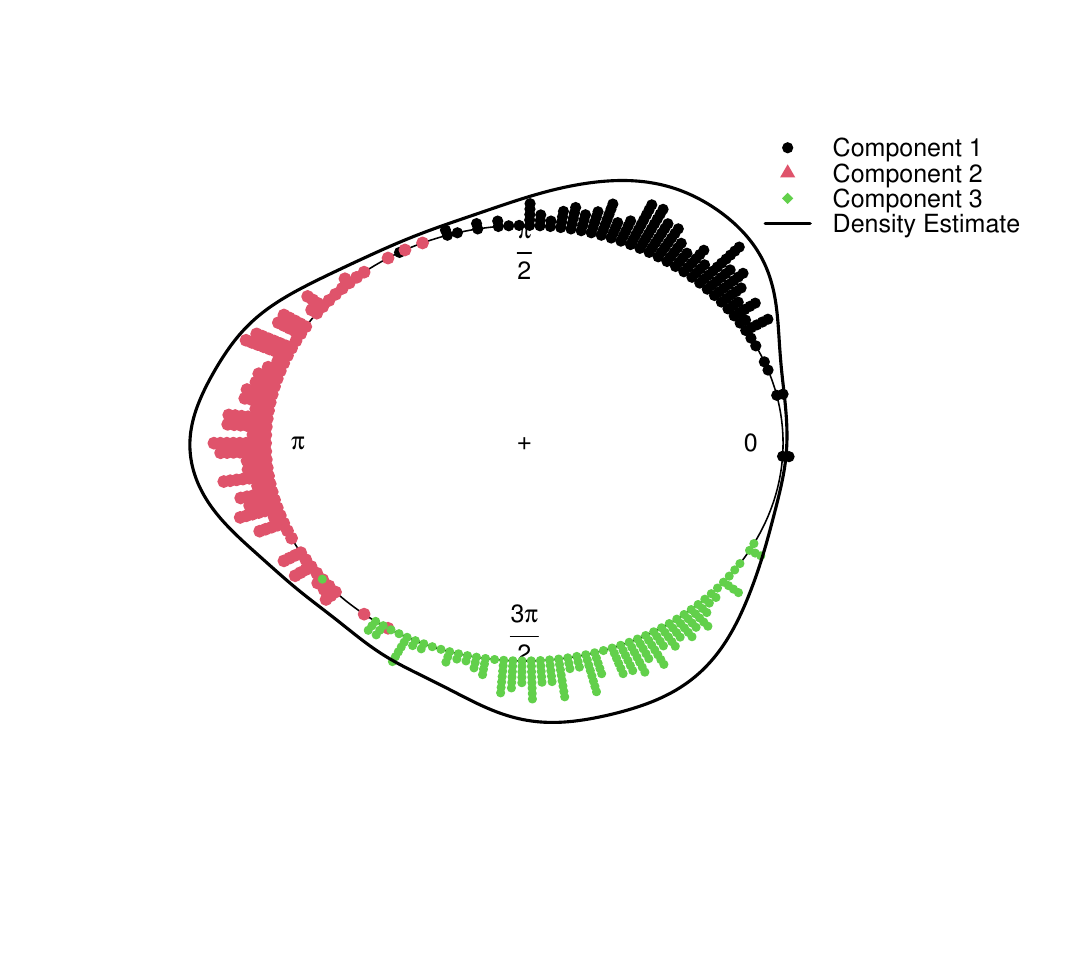}
        \caption{Non-overlapping components}
        \label{fig:plot_1_K=3}
    \end{subfigure}
    \hfill
    \begin{subfigure}{0.48\textwidth}
        \centering
        \includegraphics[width=\textwidth]{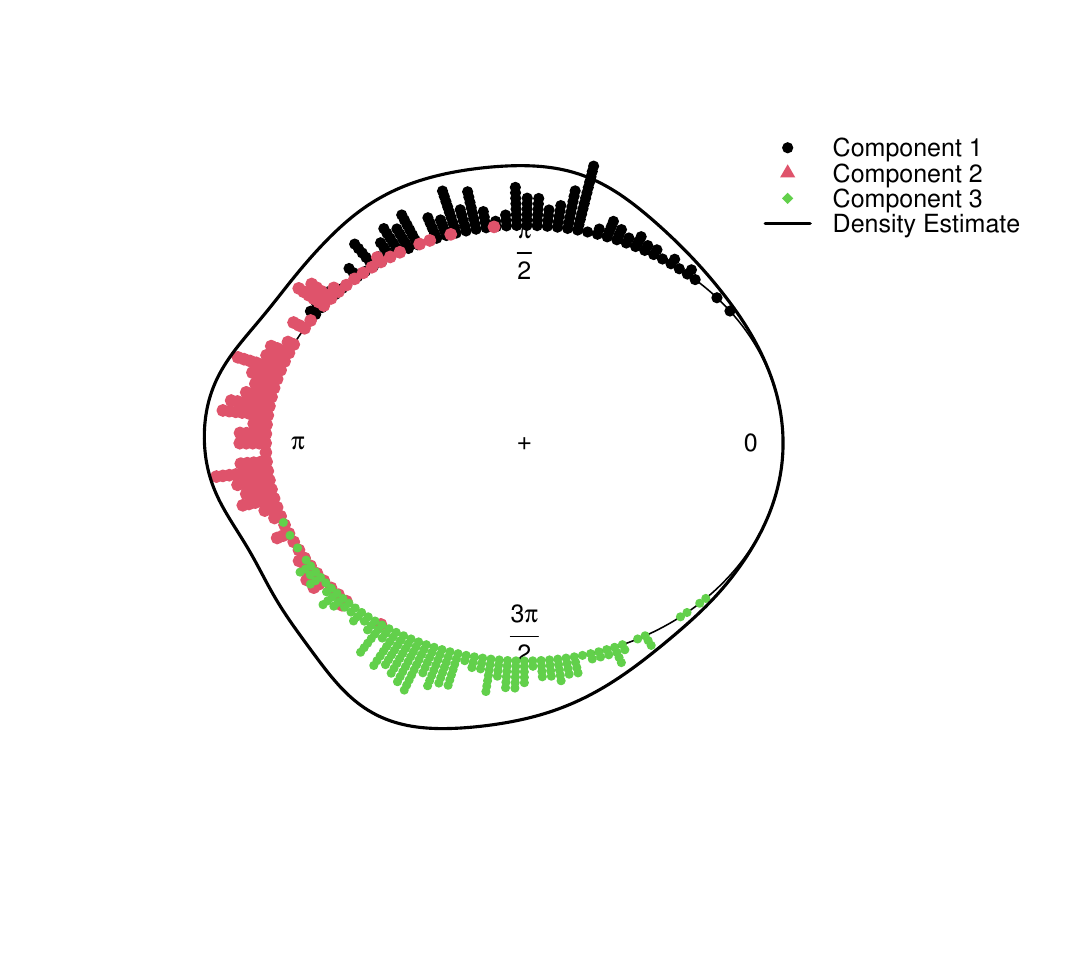}
        \caption{Overlapping components}
        \label{fig:plot_2_K=3}
    \end{subfigure}
    \caption{Circular histograms of a typical sample of size $n=500$ generated using \eqref{sim2_data} for the (a) non-overlapping and (b) overlapping case.}
    \label{fig:1}
\end{figure}

Next we demonstrate the performance of the proposed methods in detecting three non-overlapping components. The data is generated as follows
\begin{eqnarray}\label{sim2_data}
    \theta_i&=\begin{cases}
        \mathcal{VM}(\mu_1+g(\mathbf{x}_i^{*\top}\boldsymbol{B}_1),\kappa_1)\hspace{0.2cm} \text{with probability}\hspace{0.2cm} \pi_1\\
        \mathcal{VM}(\mu_2+g(\mathbf{x}_i^{*\top}\boldsymbol{B}_2),\kappa_2)\hspace{0.2cm} \text{with probability}\hspace{0.2cm} \pi_2\\
        \mathcal{VM}(\mu_3+g(\mathbf{x}_i^{*\top}\boldsymbol{B}_3),\kappa_3)\hspace{0.2cm} \text{with probability}\hspace{0.2cm} \pi_3
    \end{cases}\\
    &\text{for $i=1,2,\dots,n$}\nonumber
\end{eqnarray}
where $\mu_1=1.0996$, $\mu_2=3.1416$, $\mu_3=5.0625$, $\kappa_1=8$, $\kappa_2=6$, $\kappa_3=8$, $\pi_1=0.33$, $\pi_2=0.33$, $\pi_3=0.34$, $\boldsymbol{B}_1=(\beta_{11},\beta_{12},\beta_{13})^\top$, with $\beta_{11}=0.085$, $\beta_{12}=0.1$ and $\beta_{13}=0.3$, $\boldsymbol{B}_2=(\beta_{21},\beta_{22},\beta_{23})^\top$, with $\beta_{21}=0.09$, $\beta_{22}=0.1$ and $\beta_{23}=0.2$ and $\boldsymbol{B}_3=(\beta_{31},\beta_{32},\beta_{33})$, with $\beta_{31}=0.1$, $\beta_{32}=0.1$ and $\beta_{33}=0.1$. Figure \ref{fig:plot_1_K=3} shows a circular histogram (with a kernel density estimate overlaid on the plot) of an example of data generated using \eqref{sim1_data} for $n=500$. It is clear from Figure \ref{fig:plot_1_K=3} that the three components well-separated from each other and thus do not overlap.\\
Table \ref{tab:ex2} gives the results obtained from fitting the MixCircReg model to all the $1000$ generated samples of sizes $n=500$, $1000$ and $2000$. The results show that the proposed maximum likelihood estimators continue to exhibit good performance when the number of components are increased to $K=3$.  

\begin{table}[!ht]
\small
    \caption{RMSE of the parameter estimates for Scenario 3 and Scenario 4 over the 1000 replications}
    \centering
    \begin{tabular}{||c|ccc|ccc||}
    \hline
    Parameter&\multicolumn{3}{c}{Scenario 3}&\multicolumn{3}{|c|}{Scenario 4}\\
    \hline
    &$n=500$&$n=1000$&$n=2000$&$n=500$&$n=1000$&$n=2000$\\
    \hline
    $\pi_1$&0.0204&0.0149&0.0112&0.0228&0.0172&0.0135\\
    \hline
    $\pi_2$&0.0218&0.0149&0.0118&0.0281&0.0224&0.0146\\
    \hline
    $\pi_3$&0.0213&0.0154&0.0108&0.0279&0.0184&0.0120\\
    \hline
    $\mu_1$&0.0404&0.0292&0.0219&0.0430&0.0307&0.0247\\
    \hline
    $\mu_2$&0.0499&0.0361&0.0278&0.0508&0.0465&0.0315\\
    \hline
    $\mu_3$&0.0419&0.0308&0.0216&0.0409&0.0342&0.0256\\
    \hline
    $\kappa_1$&0.9503&0.7129&0.5062&1.0596&0.8724&0.5553\\
    \hline
    $\kappa_2$&0.8418&0.5989&0.4083&1.5114&1.1159&0.8898\\
    \hline
    $\kappa_3$&1.0043&0.7479&0.4911&1.2476&0.8287&0.5711\\
    \hline
    $\beta_{11}$&0.0322&0.0235&0.0177&0.0362&0.0257&0.0196\\
     \hline
    $\beta_{12}$&0.0483&0.0447&0.0417&0.0528&0.0444&0.0429\\
    \hline
    $\beta_{13}$&0.1277&0.1166&0.1127&0.1316&0.1161&0.1113\\
    \hline
    $\beta_{21}$&0.0389&0.0287&0.0214&0.0333&0.0327&0.0284\\
    \hline
    $\beta_{22}$&0.0544&0.0509&0.0438&0.0590&0.0531&0.0516\\
    \hline
    $\beta_{23}$&0.1359&0.1336&0.1185&0.1366&0.1225&0.1276\\
    \hline
    $\beta_{31}$&0.0342&0.0240&0.0189&0.0329&0.0287&0.0199\\
    \hline
    $\beta_{32}$&0.0502&0.0461&0.0368&0.0478&0.0473&0.0357\\
    \hline
    $\beta_{33}$&0.1271&0.1190&0.0955&0.1336&0.1180&0.0938\\
    \hline
    \end{tabular}
    \label{tab:ex2}
\end{table}
\subsubsection{Scenario 4: $K=3$ overlapping components}
We now demonstrate the performance of the proposed methods in detecting three overlapping components. The data is generated using \eqref{sim2_data} with $\mu_1=1.7279$, $\mu_2=3.1416$ and $\mu_3=4.5553$. The rest of the parameters are the same as in Scenario 3. Figure \ref{fig:plot_2_K=3} shows a circular histogram (with a kernel density estimate overlaid on the plot) of an example of a dataset generated using \eqref{sim1_data} for $n=500$. It is clear from Figure \eqref{fig:plot_2_K=2} that the three components are overlapping.\\
Table \ref{tab:ex2} gives the results obtained from fitting the MixCircReg model to all the $1000$ generated samples of sizes $n=500$, $1000$ and $2000$. The results can be interpreted similar to the results obtained in Scenario 2.\\

In summary, the proposed estimation procedure shows a good and similar performance when estimating all the model parameters. However, the model parameter estimate for $\kappa$ has a slow convergence rate compared to the other parameter estimates. This is true in all the scenarios considered. Overall, the results obtained from all the considered scenarios show that the proposed estimation procedure performs satisfactorily under various conditions. 
\subsection{Clustering performance}
The purpose of this simulation study is to evaluate the ability of the proposed model to assign observations into various homogenous groups. Mixture modelling allows partitioning of the data into clusters using the maximum a posteriori approach described in Section \ref{MBC_rule}. We will make use of the missclassification rate (ClassErr),
$\text{ClassError}=\frac{1}{n}\sum_{i=1}^n\mathbb{I}(\hat{z}_{ik}\neq z_{ik})$ and the adjusted rand Index (ARI; \cite{hubert1985}) to measure the quality of the clustering procedure. The ARI ranges between 0 and 1. This index accounts for the random assignment of data points to clusters and thus has an expected value of $0$ under random classification. We compute the average ClassErr and the average ARI over 100 replications. The ClassErr and ARI are calculated using the \texttt{classError()} function and \texttt{adjustedRandIndex()} function, respectively, from the \texttt{R} package \texttt{mclust}. The results are given in Table \ref{tab:ex4}. A small ClassErr and a large ARI are an indication that the proposed clustering procedure performs satisfactorily. The results show that this is true for both the non-overlapping cases (Scenario 1 and 3) and overlapping cases (Scenario 2 and 4) considered. However, it can be seen that the model performs slightly better in the former case compared to the latter case. This is, ofcourse, due to the difficulty in separating the components when they overlap. 
\begin{table}[ht]
  \caption{The average (and standard deviation) of the missclassification rate (ClassErr) and adjusted rand index (ARI) calculated over the 1000 replications.}
    \centering
    \begin{tabular}{||c|>{\centering}p{1.5cm}>{\centering}p{1.5cm}|>{\centering}p{1.5cm}>{\centering}p{1.5cm}|>{\centering}p{1.5cm}>{\centering}p{1.5cm}|>{\centering}p{1.5cm}p{1.5cm}||}
    \hline
        &\multicolumn{2}{c|}{Scenario 1}&\multicolumn{2}{|c|}{Scenario 2}&\multicolumn{2}{|c|}{Scenario 3}&\multicolumn{2}{|c|}{Scenario 4}\\
        \cline{2-9}        $n$&ClassErr&ARI&ClassErr&ARI&ClassErr&ARI&ClassErr&ARI\\
    \hline
$500$&0.004 (0.003)&0.985 (0.011)&0.040 (0.0103)&0.845 (0.0389)&0.012 (0.005)&0.963 (0.015)&0.0533 (0.0100)&0.850 (0.0273)\\
$1000$&0.003 (0.002)&0.987 (0.009)&0.040 (0.0066)&0.843 (0.0251)&0.012 (0.003)&0.965 (0.010)&0.0541 (0.0084)&0.850 (0.0220)\\
$2000$&0.003 (0.002)&0.986 (0.006)&0.041 (0.0054)&0.840 (0.0209)&0.012 (0.003)&0.965 (0.008)&0.0537 (0.0056)&0.850 (0.0150)\\
    \hline
    \end{tabular}
    \label{tab:ex4}
\end{table}
\section{Application}\label{applications}
\begin{figure}[h]
    \centering
    \begin{subfigure}[b]{0.45\textwidth}
    \includegraphics[width=\textwidth]{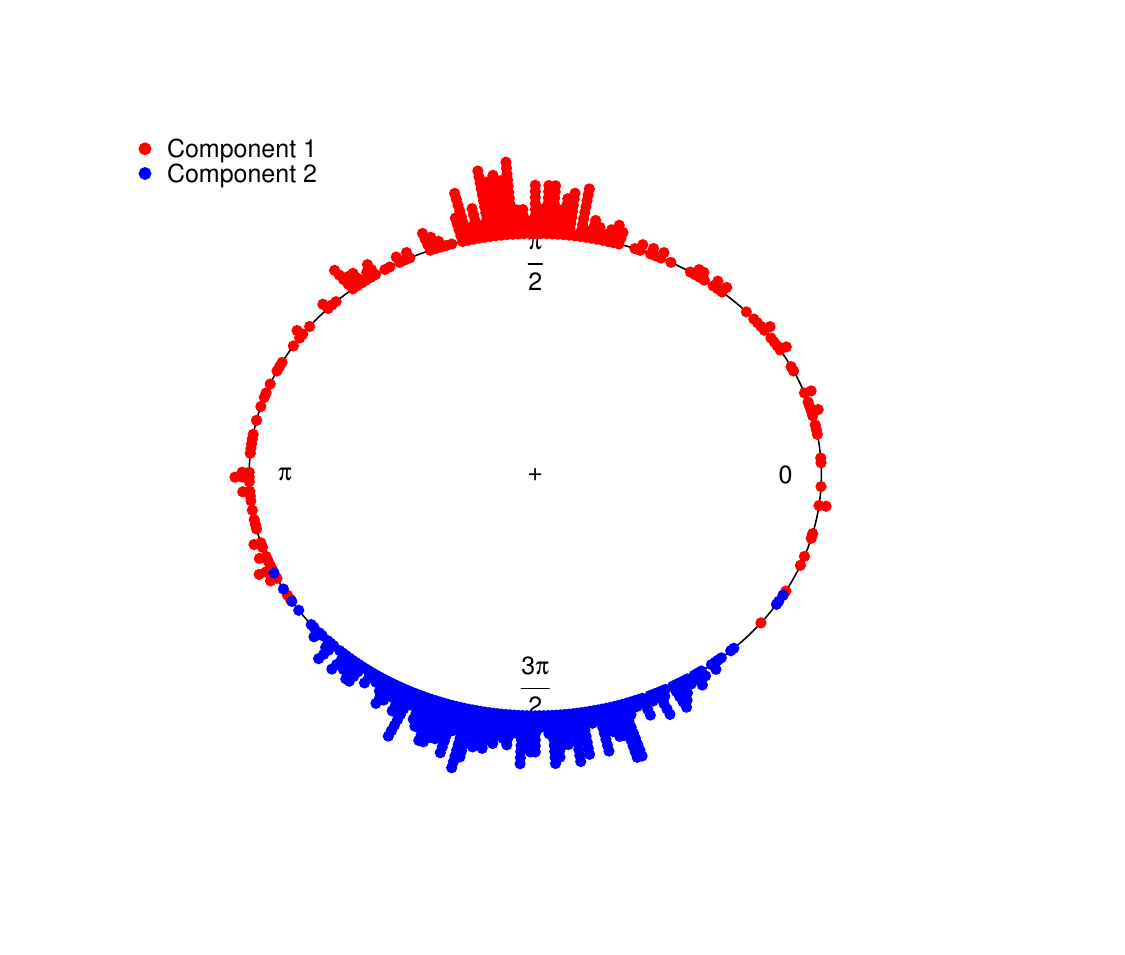}
    \caption{}
    \label{fig:est_wind}
    \end{subfigure}
    \hfill
    \begin{subfigure}[b]{0.45\textwidth}
    \includegraphics[width=\textwidth]{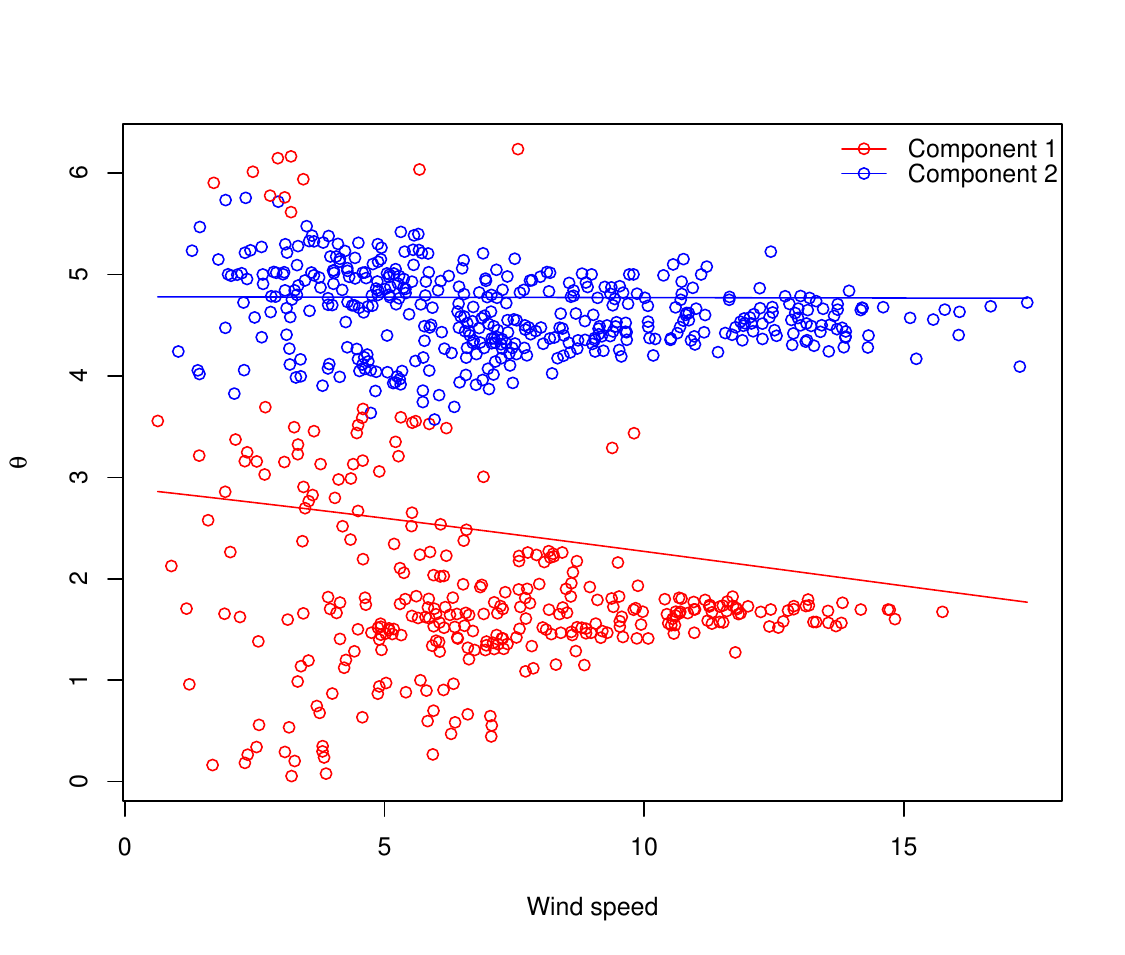}
    \caption{}
    \label{fig:scatter_est_windDS}
    \end{subfigure}
    \vfill
    \begin{subfigure}[b]{0.45\textwidth}
    \includegraphics[width=\textwidth]{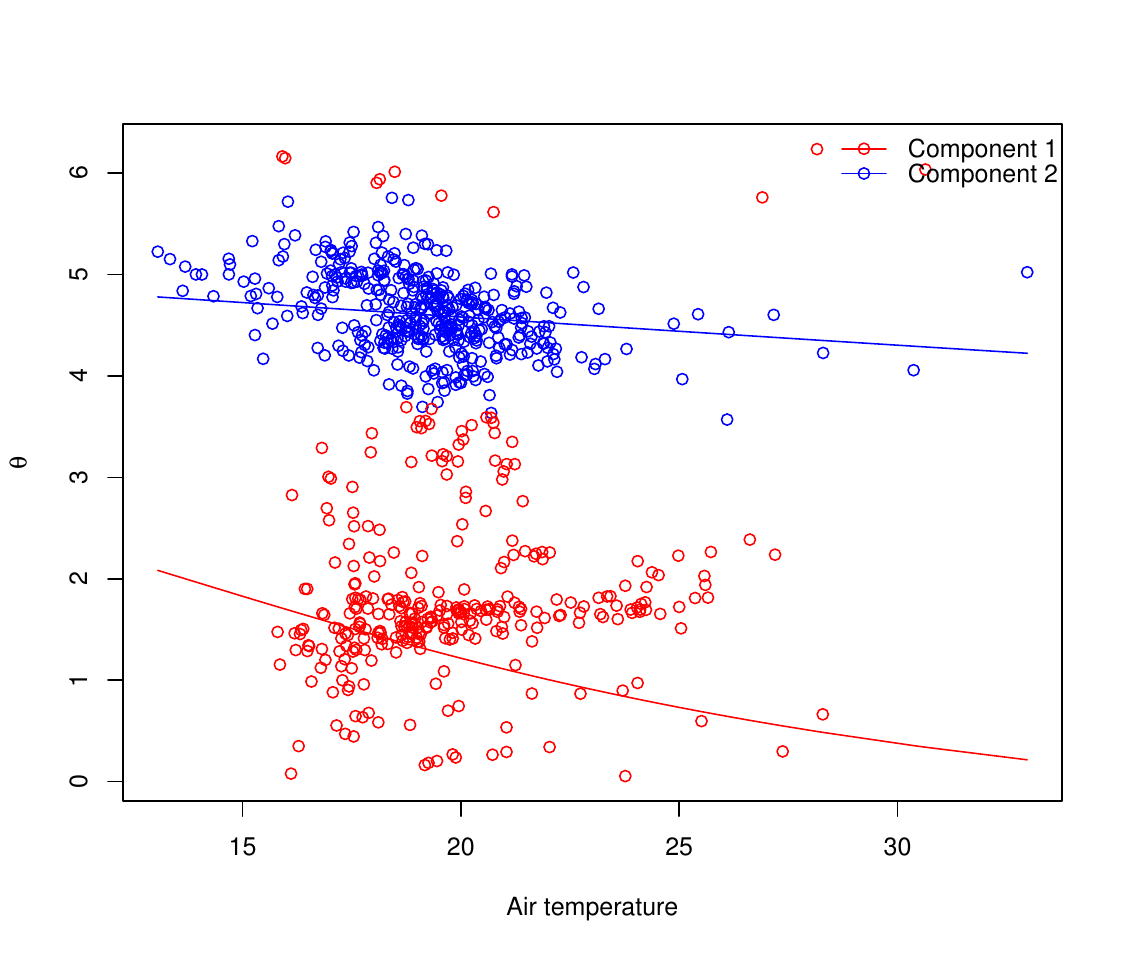}
    \caption{}
    \label{fig:scatter_est_windDA}
    \end{subfigure}
    \hfill
    \begin{subfigure}[b]{0.45\textwidth}
    \includegraphics[width=\textwidth]{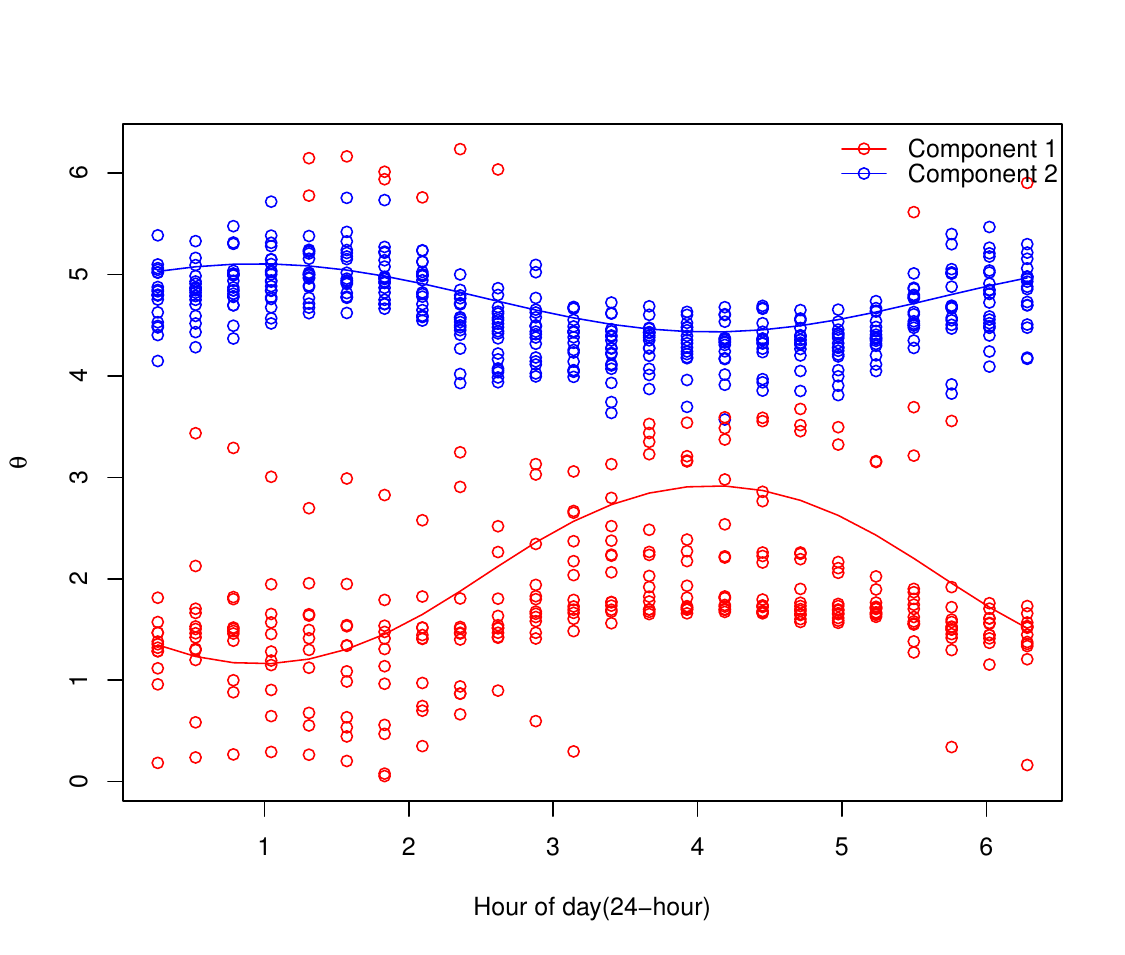}
    \caption{}
    \label{fig:scatter_est_windDH}
    \end{subfigure}
    \caption{(a) Circular plot of wind direction and (b)-(d) scatter plots of wind direction on wind speed, air temperature and hour of the day, respectively, with each point color coded based on the component it belongs to using the model-based clustering rule. The red points are associated with component 1 (east wind direction) and the blue points are associated with component 2 (west wind direction). The solid lines give the fitted regression functions for each component.}
\end{figure}

In this section, we demonstrate the practical utility of the proposed methods on the wind direction dataset introduced earlier in Section \ref{sec:intro}.\\
For wind energy production, wind direction is one of the most important measures of wind energy potential, which is the amount of wind energy that can be harnessed from a given location (\cite{rad2022}). Wind farm (location)-specific characteristics, such as wind speed and air temperature, can have an influence on wind direction. Thus, understanding the dependence of wind direction on these variables can assist with making the important decision of farm location and/or wind turbine placement in order to maximize the output of wind energy production. Our interest in this application is to use the observed wind speed, air temperature and the time (on a 24-hour clock) of day to explain the variation in the wind directions plotted in Figure \ref{fig:scatter_windDS}.\\
The wind direction data used in this application is part of a large wind direction dataset that was previously used in \cite{rad2022}, where a detailed description of the data can be found.\\
Table \ref{tab:corr} gives the correlation coefficients of the wind data. The correlation between any two linear variables is measured using the usual Pearson correlation coefficient (\cite{jammalamadaka2006}). On the other hand, the correlation between the circular variables, wind direction and time of day, is measured using the circular correlation measure in Section 8.2 of \cite{jammalamadaka2001}. Whereas the correlation between the circular variable, wind direction, and the linear variables, wind speed and air temperature, is measured using the correlation measure in Section 8.5 of \cite{jammalamadaka2001}.\\
As already mentioned in the introduction, the observed wind direction is bimodal (see Figure \ref{fig:circ_wind}). Thus, we fit the $K=2-$component model
\begin{eqnarray}\label{model_wind}
\mathcal{MVM}(\theta|\mathbf{X}^\ast=\mathbf{x}^\ast)=\pi_1\mathcal{VM}(\mu_1+g(\mathbf{x}^\ast\boldsymbol{\beta}_1),\kappa_1)+\pi_2 \mathcal{VM}(\mu_2+g(\mathbf{x}^\ast\boldsymbol{\beta}_2),\kappa_2),
\end{eqnarray}
where $\theta=$ wind direction, $\phi=$ hour of the day and $\mathbf{x}^\ast=(\mathbf{x}_1,\mathbf{x}_2)^\intercal$, where $\mathbf{x}_1=(\text{sin}(\phi),\text{cos}(\phi))^\intercal$ and $\mathbf{x}_2=(\text{Wind Speed},\text{Air temperature})^\intercal$. The parameter estimates of the fitted model are given in Table \ref{tab:est_wind}. We also calculated the standard errors and 95\% confidence intervals using the following parametric bootstrap approach:
\begin{enumerate}
    \item For a given value of $x$, we sample the corresponding value of the response, denoted by $\theta^\ast$, from the fitted model (\ref{model_wind}), $\hat{\pi}_1\mathcal{VM}(\hat{\mu}_1+g(x\hat{\beta}_1),\hat{\kappa}_1)+\hat{\pi}_2 \mathcal{VM}(\hat{\mu}_2+g(x\hat{\beta}_2),\hat{\kappa}_2)$. We repeat this sampling process $n$ times to get a bootstrap sample $\mathcal{S}=\{(x_i,\theta_i^\ast):i=1,2,\dots,n\}$. We generate $B$ bootstrap samples $\mathcal{S}^{(1)},\mathcal{S}^{(2)},\dots,\mathcal{S}^{(B)}$ in the above manner.
    \item Next,  We fit model (\ref{model_wind}) on each of these bootstrap samples, thus generating a sampling distribution of $\hat{\pi}_k$, $\hat{\mu}_k$, $\hat{\kappa}_k$ and $\hat{\beta}_k$, for $k=1,2,\dots,K$.
    \item Finally, to compute the 95\% confidence intervals, we take the $2.5^{th}$ and $97.5^{th}$ percentiles of the sampling distributions as the lower and upper limits, respectively, of the interval. We set $B=1000$.
\end{enumerate}
We can see from Table \ref{tab:est_wind} that for the majority of the time, about $57\%$, the wind came from the north-west direction, which is estimated as about $4.8$ (in radians) or $275\textdegree$. All the estimated regression coefficients are statistically significant except for the coefficient of wind speed when the wind comes from the north-west (second component with estimated wind direction of about $275\textdegree$).\\ 
We also fitted the circular regression model ($K=1$) and obtained a BIC of $2586.414$ compared with a BIC of $1787$ for the fitted model \eqref{model_wind}. This shows that, compared to the latter model, model \eqref{model_wind} gives the best fit for the wind data.\\
Lastly, to demonstrate the clustering capability of the fitted model, we can use the model-based clustering rule to assign each observation to a given wind direction. 
Figures \ref{fig:est_wind}-\ref{fig:scatter_est_windDH} show the circular plot and the scatter plots of wind direction to wind speed, air temperature and hour of the day, respectively. The points are color coded based on the component they have been classified into using the model-based clustering rule described in Section \ref{MBC_rule}. Also included in the scatter plots, are the estimated component circular regression functions. 

\begin{table}[h]
    \caption{Correlation coefficients of the wind data.}
    \centering
    \begin{tabular}{||p{3cm}|p{2.5cm}|p{2.5cm}|p{2.5cm}|p{2.2cm}||}
    \hline
        &Wind direction&Wind speed&Air Temperature&Time of day (hour)\\
        \hline
         Wind direction&1&-&-&-\\
         Wind speed&0.0454$^\ast$&1&-&-\\
         Air Temperature&0.0390$^\ast$&0.0505&1&-\\
         Time of day (hour)&-0.4076$^\ast$&0.0451$^\ast$&0.1800$^\ast$&1\\
         \hline
          \multicolumn{5}{|l|}{*Significant at 0.001}\\
         \hline
    \end{tabular}
    \label{tab:corr}
\end{table}

\begin{table}[ht]
    \caption{Parameter estimates (with standard errors (std error)), 95\% confidence intervals and fit statistics for model (\ref{model_wind}) fitted on the wind data. The standard errors and confidence intervals were obtained via the proposed parametric bootstrap procedure.}
    \centering
    \begin{tabular}{||c|c|c||}
    \hline
         Parameter&Estimate (std error)&95\% Bootstrap CI\\
    \hline
         $\pi_1$&0.4294 (0.0180)&$(0.3934,0.4645)$\\
         $\mu_1$&2.0388 (0.1382)&$(1.7614,2.3150)$\\
         $\mu_2$&4.7690 (0.0497)&$(4.6730,4.8700)$\\
         $\kappa_1$&3.0825 (0.2516)&$(2.6640,3,6167)$\\
         $\kappa_2$&13.4460 (1.0343)&$(11.7314,15.7270)$\\
         \cline{2-3}
         \multirow{4}{*}{$\boldsymbol{\beta}_1$}&-0.2695 (0.0367)& (-0.3448,-0.2046)\\
         &-0.3838 (0.0376)&(-0.4619,-0.3164)\\
         &-0.0679 (0.0417)&(-0.0921,-0.0456)\\
         &-0.1897 (0.0842)&(-0.2702,-0.1270)\\
         \cline{2-3}
         \multirow{4}{*}{$\boldsymbol{\beta}_2$}&0.0991 (0.0111)&(0.0775,0.1211)\\
         &0.1382 (0.0115)&(0.1161,0.1604)\\
         &-0.0009 (0.0152))&(-0.009,0.009)\\
         &-0.0406 (0.0240)&(-0.0617,-0.0196)\\
         \hline\hline
        $\ell({\hat{\vartheta}})$&\multicolumn{2}{|c||}{-851}\\
         BIC&\multicolumn{2}{|c||}{1787}\\
         \hline
    \end{tabular}
    \label{tab:est_wind}
\end{table}
\section{Conclusion}\label{discussion}
In this paper, we proposed a mixture of circular regression models to study the dependence of a circular response on a set of linear (or real-valued) and/or circular covariates whenever the data are multimodal. To estimate this model, we proposed maximum likelihood estimates obtained using the Expectation-Maximization algorithm. The performance of the model and estimation procedure was demonstrated through Monte Carlo simulations. Moreover, to demonstrate the practical utility of the proposed model, we applied it to a real wind dataset to study the influence of wind speed, air temperature and the time of day on wind direction. We demonstrated the advantage of the fitted model by using it to classify the observed wind directions into the two clusters identified by the model.\\
For future studies, it would be interesting to extend the proposed model to accommodate outliers (or atypical observations) and skewness in the circular response variable. 

\section*{Declarations}

\paragraph{Competing interests:} The authors have no competing interests to declare that are relevant to the content of this article.
\paragraph{Data availability:} The data and code used in this paper is available through the link: \url{https://github.com/Sphiwe-Skhosana/MixCircReg}

\bibliography{References}
\bibliographystyle{plainnat}
\end{document}